\documentclass[iop,apjl,letterpaper]{emulateapj}
\usepackage{natbib}
\usepackage{amsmath}
\usepackage{url}

\newcommand       \be		{\begin{equation}}
\newcommand       \ee		{\end{equation}}
\newcommand       \bea          {\begin{eqnarray}}
\newcommand       \eea          {\end{eqnarray}}

\newcommand       \pc		{\,{\rm pc}}

\newcommand       \yr		{\,{\rm yr \,}}
\newcommand       \Myr		{\,{\rm Myr \,}}
\newcommand       \cm		{\,{\rm cm \,}}
\newcommand       \s		{\,{\rm s \,}}
\newcommand       \g		{\,{\rm g \,}}
\newcommand       \kms          {\,{\rm km \,\, s}^{-1}}
\newcommand       \eff          {\epsilon_{\rm ff}}
\newcommand       \tff          {\tau_{\rm ff}}
\newcommand       \tdyn         {\tau_{\rm dyn}}

\begin{document}

\title{TIME-VARYING DYNAMICAL STAR FORMATION RATE}
\author{Eve J. Lee\altaffilmark{1,2}, Philip Chang\altaffilmark{1,3},
  Norman Murray\altaffilmark{1,4}} 
\altaffiltext{1}{Canadian Institute for Theoretical Astrophysics, 60
  St.~George Street, University of Toronto, Toronto ON M5S 3H8, Canada} 
\altaffiltext{2}{Astronomy Department, University of California,
  Berkeley CA 94720, USA; evelee@berkeley.edu} 
\altaffiltext{3}{Department of Physics, University of
  Wisconsin-Milwaukee, 1900 E. Kenwood Blvd., Milwaukee, WI 53211, USA} 
\altaffiltext{4}{Canada Research Chair in Astrophysics}

\begin{abstract}
We present numerical evidence of dynamic star formation in which the
accreted stellar mass grows superlinearly with time, roughly as
$t^2$. We perform simulations of star formation in
self-gravitating hydrodynamic and magnetohydrodynamic turbulence
that is continuously driven. By turning the self-gravity of the gas in
the simulations on or off, we demonstrate that self-gravity is the
dominant physical effect setting the mass accretion rate at early
times before feedback effects take over, contrary to theories of
turbulence-regulated star formation. We find that gravitational
collapse steepens the density profile around stars, generating the
power-law tail on what is otherwise a lognormal density probability distribution
function. Furthermore, we find turbulent velocity profiles to flatten
inside collapsing regions, altering the size-linewidth relation. This
local flattening reflects enhancements of turbulent velocity on small
scales, as verified by changes to the velocity power spectra. Our
results indicate that gas self-gravity dynamically alters both density
and velocity structures in clouds, giving rise to a time-varying star
formation rate. We find that a substantial fraction of the gas that
forms stars arrives via low-density flows, as opposed to accreting
through high-density filaments.
\end{abstract}

\section{Introduction}
\label{sec:intro}
Star formation in galaxies proceeds at a leisurely pace: the time to
turn all the molecular gas into stars is much longer than the disk dynamical
time \citep{1998ApJ...498..541K,2008AJ....136.2782L}. Denoting the
dynamical time by $\tau_{\rm dyn}\equiv R_d/v_c$, where $R_d$ is the
half-light radius of the disk and $v_c$ is the circular velocity of
the galaxy, the star formation rate (SFR) is observed to be 
\be
\dot M_*=\eta
M_g/\tau_{\rm dyn}, 
\ee
with $\eta\approx 0.017$. In other words, were
new supplies not available to the disk, it would take roughly 50
dynamical times to deplete the gas.

One can envision extending the relation between the SFR, gas mass, and 
dynamical time to smaller
scales, including giant molecular clouds (GMCs) or star-forming clumps:
\be \label{eqn: slow}
\dot M_* = \eff M_g/\tff, 
\ee
where $\dot M_*$ now refers to the SFR in the host
GMC (or a smaller feature such as a clump), $M_g$ is the associated
total mass (initially all gas), $\tff$ is the free-fall time of the
GMC or clump, and $\eff$ is a dimensionless number referred
to as the SFR per free fall time, analogous to
$\eta$. The SFR is not to be confused with the star
formation efficiency SFE~$\equiv M_*/(M_* + M_g)$, where $M_*$ is the total
stellar mass. The free-fall time is defined as
\be  \label{eqn: tff}
\tff\equiv\sqrt{3\pi\over 32 G\bar\rho},
\ee
where $\bar\rho$ is the mean density in the region of interest.

\citet{zuckerman74} first realized that should all the Galactic
molecular gas ($\sim$$10^9 M_\odot$) collapse in a
free-fall time, the expected SFR would be approximately two orders of
magnitude higher than the observed Galactic SFR, the observed rate
corresponding to $\eff$$\sim$0.02.\footnote{The number
  they quote is $\eff$$\sim$0.13 but they underestimate the total mass of
  Galactic molecular gas.
Because the correct value is $\eff \simeq 0.02$, $\eff$ has often been 
taken as a constant
$\sim0.02$~\citep[e.g.,][]{2005ApJ...630..250K,2012ApJ...745...69K}.}
Based on the assumption of a static density probability distribution
function (PDF) of lognormal form, \citet{2005ApJ...630..250K} predict
$\eff\sim0.02$ with little scatter. The theory consists of choosing a
critical density $\rho_{\rm crit}$ above which gas is believed to collapse
into stars, and integrating up the density PDF from $\rho_{\rm crit}$ to
infinity, resulting in a stellar mass. This stellar mass is then
divided by the volume-averaged free-fall time. 

There are reasons to question this simple result and the constancy of
$\eff$. Recent numerical studies
\citep[e.g.,][]{klessen00,dib05,vazquez08,2008PhST..132a4025F,ballesteros11,cho11,2011ApJ...727L..20K,2012ApJ...750...13C,2013ApJ...763...51F}
show the emergence of a power-law tail at the high-density end of the density PDF
due to gravitationally induced collapse. The idea that gravitational collapse generates 
high-density tail is bolstered by the observations of extended tails in the column density PDFs 
in actively star-forming clouds, in contrast to the lognormal PDFs
seen in non-star-forming clouds 
\citep[e.g.,][]{kainulainen09, schneider13}.
This raises concerns for any theory of star
formation that employs a static log-normal density PDF.\footnote{
  An analytic model to explain the evolution of a power-law density tail
  has been proposed \citep{girichidis14} but the authors assume
  pressure-free collapse. We show in this paper (and further in
  \citealt{murray14}) that there is a close interplay between gravity
  and turbulence whose pressure input cannot be ignored.} 
Furthermore, observations suggest that there is a wide range in
$\eff$. \citet{1988ApJ...334L..51M} find a range in excess of a factor
of 100 in the ratio of far-infrared flux (a proxy for the mass
of young stars) to CO line flux in a
sample of molecular clouds in the Milky Way. Using different probes of
the SFR such as counts of
protostellar objects \citep{2010ApJ...723.1019H, 2010ApJ...724..687L},
infrared luminosities of massive clumps traced in HCN
\citep{2010ApJS..188..313W}, and free-free emission and far-infrared emission
of massive star-forming regions \citep[][where the former authors use
free-free emission and hydrogen recombination lines to estimate the
stellar mass of OB associations]{1997ApJ...476..166W,2011ApJ...729..133M}, other
authors find a similar range in $\eff$, spanning at least two decades.

Using the data in \citet{2010ApJ...723.1019H} and
\citet{2010ApJ...724..687L}, \citet{2012ApJ...745...69K} arrive at a
different conclusion than the authors of the former two papers.
\citet{2012ApJ...745...69K} argue that SFR on GMC
($\sim 100\pc$) and smaller scales is as slow as the galaxy-wide 
SFR with a range in $\eff$ of order only a factor of 10.
They reason that the spread in $\eff$ shrinks once different
free-fall times are taken into account. \citet{2012ApJ...745...69K}
conclude that star formation is universally slow at all times and
at all scales. The data they consider, however, only pertain to nearby
star-forming clouds. 

In contrast, \citet{federrath13} considers a larger 
sample of local clouds and simulation results to find factors of $\sim$100 
scatter around constant $\eff$$\sim$2\%, using the same method as 
\citet{2012ApJ...745...69K} to take different free-fall times into account.

Simulations suggest that turbulence does not limit
the rate of star formation to $\eff\sim0.02$
\citep[e.g.,][]{2010ApJ...709...27W,cho11,2011ApJ...730...40P,2012MNRAS.419.3115B,2012ApJ...754...71K,2012ApJ...761..156F,2014MNRAS.439.3420M}.
\citet{2011ApJ...730...40P}
present both hydrodynamic (HD) and magnetohydrodynamic (MHD)
simulations of star formation in turbulently stirred isothermal
gas. They find rapid star formation, $\eff\sim0.5-0.9$, in simulations
with a virial parameter between $0.5$ and $1$; including the effects
of magnetic fields, they find a decrease in the simulated SFR of 
about a factor of three: $\eff\sim0.2$--$0.4$. 
\citet{2012ApJ...754...71K} include stellar wind in
HD simulations and find $\eff\sim0.3$. 
\citet{2014MNRAS.439.3420M}
include both stellar wind and magnetic fields to find $\eff\sim0.1$. 
When protostellar outflows are important---typically in regions 
of low-mass star formation---SFE is usually reduced only by factors 
of $\sim$3 \citep{matzner00,hansen12,machida13,offner14,federrath14}.

In this paper we conduct large-scale (16 pc) HD and MHD simulations of
star-forming clouds with continuously driven supersonic
turbulence. Like previous simulations, we will find large values of
$\eff \sim 0.3$. But we will also discover that $\eff$ is not constant
with time---we will find that $\eff \propto t$. In fact, this time dependence is evident
in the aforementioned simulations (e.g., Figures 5 and 6 of \citealt{2011ApJ...730...40P};
Figure 5 of \citealt{2012ApJ...754...71K} ; Figure 7 of
\citealt{2014MNRAS.439.3420M}). 
All these simulations (including ours) are characterized by an initial
lull in SFE. This feature was commonly considered
a transient borne out of the artificial and sudden inclusion of gas
self-gravity. We argue that this interpretation is not correct: the
initially slow rate and low efficiency of star formation
are direct consequences of the interplay of self-gravity and turbulent
pressure. The physical significance of the initial quiescence in SFE was
recognized by \citet{cho11} who also find a time-varying $\eff$ and note the 
role of self-gravity in shaping the turbulent structure of the gas.
We will present quantitative analysis to support this assertion.
\citet{2014MNRAS.439.3420M} find that $\eff$ is
time dependent in their HD and MHD simulations with decaying
turbulence---they further speculate that continuously driven
turbulence should suppress the time dependency of
$\eff$. We will show explicitly using continuously driven turbulent
simulations that this is not the case.

This paper is the first in a series of four showing that on GMC
scales, $\eff$ is neither constant
nor small. We find that in gravitationally bound clouds, the 
SFR increases with time and can reach $\eff\sim0.3$ or
higher. In this paper (Paper I), we present the evidence from our
numerical simulations while in Paper II \citep{murray14},
we present an analytical
model based on adiabatically heated turbulence 
\citep{2012ApJ...750L..31R} to explain the results of Paper I. 
In Paper III (E.J. Lee et al., in preparation), we
present observational evidence for dynamical star formation
using the most complete census of GMCs in the Milky Way to date, and 
we find that $\eff$ ranges from $\sim$$10^{-4}$ to $\sim$$0.4$. 
The GMC catalog and the details of its construction are contained
in Paper IV (M-A. Miville-Desch\^{e}nes et al., in preparation).

We follow the practice of others in this field in using the formation
and growth of sink particles as a proxy for star formation \citep[e.g.,]
{2011ApJ...730...40P,
  2012ApJ...754...71K,2013ApJ...763...51F}. This proxy is not exact,
since stars, which are hydrostatic objects, form at much higher
density and on much smaller scales (of order $10^{12}\cm$ at most)
than are simulated in our simulations or in those just cited---the
dynamic range in these simulations is typically no larger than 1000,
so in a 1 pc box the smallest resolved scale is at least 1000 times
larger than a star. Recent simulations using adaptive mesh refinement
have achieved a dynamic range of eight orders of magnitude, e.g.,
\citet{2013ApJ...763....6T}, large enough to resolve the outer layers
of a protostar. However, such simulations run for only hundreds to
thousands of years, not nearly long enough to determine an SFR. 
The reason, of course, is that the dynamical or sound
crossing times on the smallest scale are so short.

This paper is organized as follows. In Section \ref{sec:numsim} we 
briefly describe our numerical
methods. In Section \ref{sec:analy} we analyze the results of our
simulations. We discuss our results and compare them to previous work
in Section \ref{sec:discussion}, and we conclude in Section \ref{sec:concl}.

\begin{figure}[!htb]
\plotone{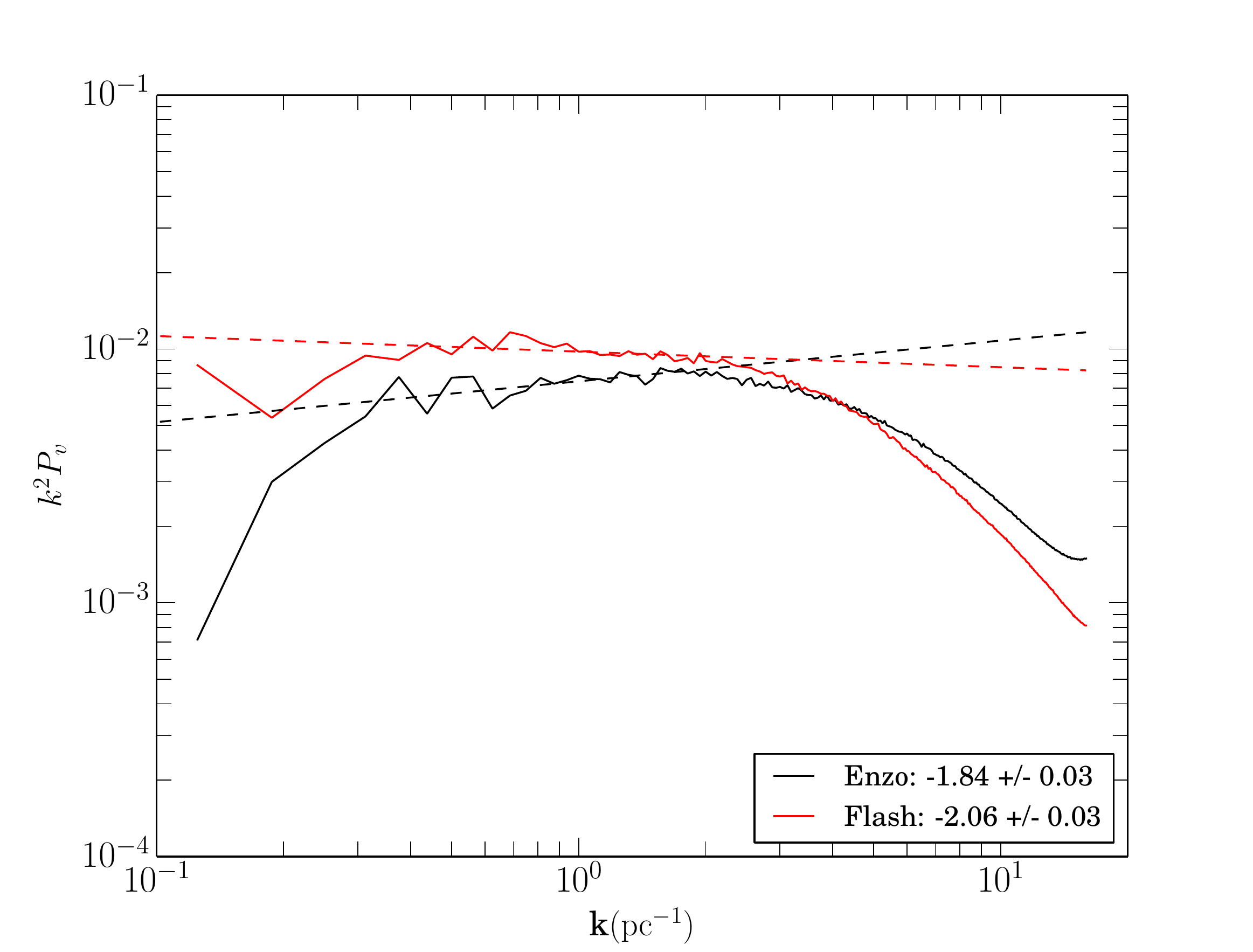}
\caption{\label{fig:turb_b4grav}Velocity power spectrum, compensated by
  $k^2$, of the fully turbulent box in the absence of self-gravity. The
  solid black curve is the power spectrum of the rms
  velocity in ENZO while the red curve is for FLASH. The dashed line is the 
  power-law fit to the
  inertial range where the power-law index and its error are given in
  the legend. The results when self-gravity is turned on (not shown) are very
  nearly the same. The striking difference between the two codes at smaller $k$ 
  is due to different driving scales; ENZO is driven at $3 \leq kL \leq 4$, as opposed to 
  FLASH's $1 \leq kL \leq 2$.}
\end{figure}

\section{NUMERICAL METHODS}
\label{sec:numsim}

We use both ENZO (v2.2 development branch; \citealt{ENZO}) and FLASH
4.0.1 to perform our numerical calculations. 
Hydrodynamical runs solve the continuity and 
momentum equations, 
with Poisson's equation when self-gravity is included:
\be
{\partial\rho\over \partial t} + \nabla\cdot (\rho {\bf v})=0,
\ee
\be 
{\partial \rho {\bf v}\over \partial t} + \nabla\cdot
\left(
\rho{\bf vv} - {1\over 4\pi}{\bf BB}+{\bf I}(p + B^2/8\pi)
\right) 
= -\rho\nabla\phi,
\ee 
and 
\be 
\nabla^2\phi=4\pi G\rho.
\ee 
In these equations ${\bf v}$ is the fluid velocity, ${\bf B}$ is the
magnetic field, ${\bf I}$ is the identity matrix, and $\phi$ is the gravitational potential. The gas
pressure is given by
\be 
p=\rho c_s^2,
\ee 
where the sound speed $c_s$ is taken to be constant. The evolution of
the magnetic field is given by
\be 
{\partial {\bf B}\over\partial t} + \nabla\times({\bf B}\times{\bf v})=0.
\ee 
Readers interested in the architecture of 
ENZO should refer to \citet{ENZO}; the relevant equations are 
their (1)--(8) with $a=1$, $\dot{a}=0$ (non-cosmological), 
$\Lambda=\Gamma=F_{\rm cond}=0$ (no radiation, no heating, no conduction).
Our ENZO runs use the Runge
Kutta second-order based MUSCL solver with a Harten-Lax-van Leer (HLL)
Riemann solver. The FLASH runs  use the unsplit solver recently
developed by \citet{2009JCoPh.228..952L}, with 
the Harten-Lax-van Leer
Contact (HLLC) Riemann solver for the HD case and the
the Harten-Lax-van Leer with
Discontinuities (HLLD) Riemann solver for the MHD case. We note that the HLL solver is, by construction, 
more diffusive and does not resolve shocked
regions as well as the HLLC solver. While the two codes disagree at a
factor of two level on subparsec scales, the global results are in good
agreement.

Our simulations start with uniform density and zero bulk velocity 
  with a fixed resolution of $512^3$. In our fiducial runs, the
physical length of the box is set at $L=16\pc$, and we use periodic
boundary conditions. The initial mass density is
$3\times10^{-22}\g\cm^{-3}$. The global free-fall time is then
$\tff\approx3.9\Myr$. As mentioned above, all our simulations are
isothermal at a sound speed $c_{s} = 2.65\times10^4 \cm\s^{-1}$.  
  Thermodynamic studies find that molecular clouds can contain
  significant amount of atomic gas with temperatures ranging from that
  of cold neutral medium (CNM) and warm neutral medium (WNM), on the length scales
  and densities similar to our setup
  \citep[e.g.,][]{koyama02,audit05,heitsch08,banerjee09,audit10,vazquez10,heiner14}.
  Clouds that harbour star-forming clumps may well have temperature
  variations but as \citet{heitsch08} report, the conversion from
  atomic to molecular hydrogen is a rapid process, and stars form only
  in the gravitationally bound molecular clumps. Given the same
  density threshold for star formation, \citet{heitsch08} find
  $\epsilon_{\rm ff} \lesssim 0.5$, similar to what we find (see their
  Figure 10) in our isothermal models.

In our MHD runs, we set the initial magnetic field to be uniform, 
  pointing in the x-direction, and to have an intensity of
0.49$\mu$G, yielding an initial $\beta_{B} \equiv 2c_{s}^{2}/v_{A}^{2}
\sim 22.2$, where $v_{A}$ is the Alfv\'{e}n speed. At statistical
  equilibrium, $\beta_B \sim 1.3$, corresponding to $|B| \sim 2\mu G$,
  on the low side but still within the range of observed magnetic
  field intensity of molecular clouds of similar
  densities~\citep[see][their Figure 6]{crutcher12}.  Our simulated
  box is magnetically supercritical ($M/M_\Phi \sim 20$ where $M_\Phi
  = 0.12\pi B (L/2)^2/\sqrt{G}$) and so are GMCs. For the typical
  turbulent velocities of Milky Way GMCs $v_T \sim 7 \kms$ and size
  100 pc, number densities of $n_{\rm H} \sim 100 \cm^{-3}$ are
  required for virial equilibrium. The maximum possible magnetic field
  intensity is then $\sim$$10 \mu G$ \citep{crutcher12}, corresponding
  to $M/M_\Phi \sim 10$.

Initially, we evolve the gas without any self-gravity and drive
turbulent forcing in Fourier space at wavenumbers $1 \leq kL \leq 2$
($3 \leq kL \leq 4$ for ENZO) until the gas becomes fully
turbulent at the appropriate Mach number ${\cal M}\equiv v_{\rm
  T,0}/c_s = 9$, where $v_{\rm T,0}$ is the turbulent velocity at the
largest scale $L/2$. In most of our runs, with either code, we employ a
  purely solenoidal form of turbulent driving. We have also done some
  runs employing compressive driving. 
  Compressive driving has been shown, e.g., by
  \citet{2012ApJ...761..156F}, to lead to much larger SFRs than does 
	solenoidal driving. We have concentrated on the more conservative
  form of driving to reduce the amount by which we may be
  overestimating the SFR, but we also report on the
  results of simulations with compressive driving.

Even in cases where we drive the turbulence on large scales with
  purely solenoidal driving, on smaller scales the turbulence has both
  solenoidal and compressive modes. In runs with purely solenoidal driving
  (using FLASH), the ratio between the
  turbulent power due to solenoidal and compressive modes is 5:1 at
  the driving scale, converging to 1:1 at the dissipation scale, i.e.,
  the point where $k^2P_v(k)$ deviates from the flat spectrum seen in
  Figure \ref{fig:turb_b4grav}, at $k\approx 30\,\pc^{-1}$. Similar power 
	ratios have been observed in previous studies such as \citet{federrath10aa} 
	(see their Figure 14). 
	In our ENZO runs, the power in the compressive mode becomes comparable to
  that in the solenoidal mode only at scales a few times the grid
  scale. The difference is likely due to some combination of the
  different driving scales (ENZO's driver drives at slightly smaller
  scales, as noted above) and ENZO's more dissipative solver.

Away from collapsing regions, the turbulence is found to follow the
size-linewidth relation

\be \label{eqn: size-linewidth}
v_{\rm T}(r)=v_{\rm T,0}(2r/L)^p,
\ee

\noindent where $p\approx0.5$, consistent with Burgers turbulence
\citep{burgers}, i.e., high Mach number turbulence, and as
  observed in GMCs and low-mass star-forming regions by
\citet{1988ApJ...329..392M}. The dynamical time is $\tdyn\equiv
L/2{\cal M}c_s\approx3(9/{\cal M})\Myr$. The simulation reaches a
statistical steady state after 3--5 $\tau_{\rm dyn}$, whereupon the
global virial parameter
\be
\label{eqn: avir}
\alpha_{\rm vir}\equiv 5{v_{\rm T,0}^2 (L/2)\over 3GM}
\ee
\noindent reaches, by design, unity. In this expression $M$ is the
total mass in the box.

The velocity power spectrum of the fully turbulent state is shown in Figure
\ref{fig:turb_b4grav}.  The measured index of this turbulence
($\sim$1.8--2.1) is close to the expected value for Burgers
turbulence and consistent with the value of $p\approx 1/2$ found above.

Both the sonic length
\be
l_s\equiv {L\over 2}\left(\frac{c_s}{v_{\rm
      T,0}}\right)^2\approx0.1\pc
\ee
and the Jeans length
\be
\lambda_J\equiv\sqrt{\pi c_s^2\over G\bar\rho}\approx3.4\pc
\ee
are well resolved in our $512^3$ runs, which have a cell length
of $3\times10^{-2}\pc$.

\begin{figure*}[!tb]
 \plottwo{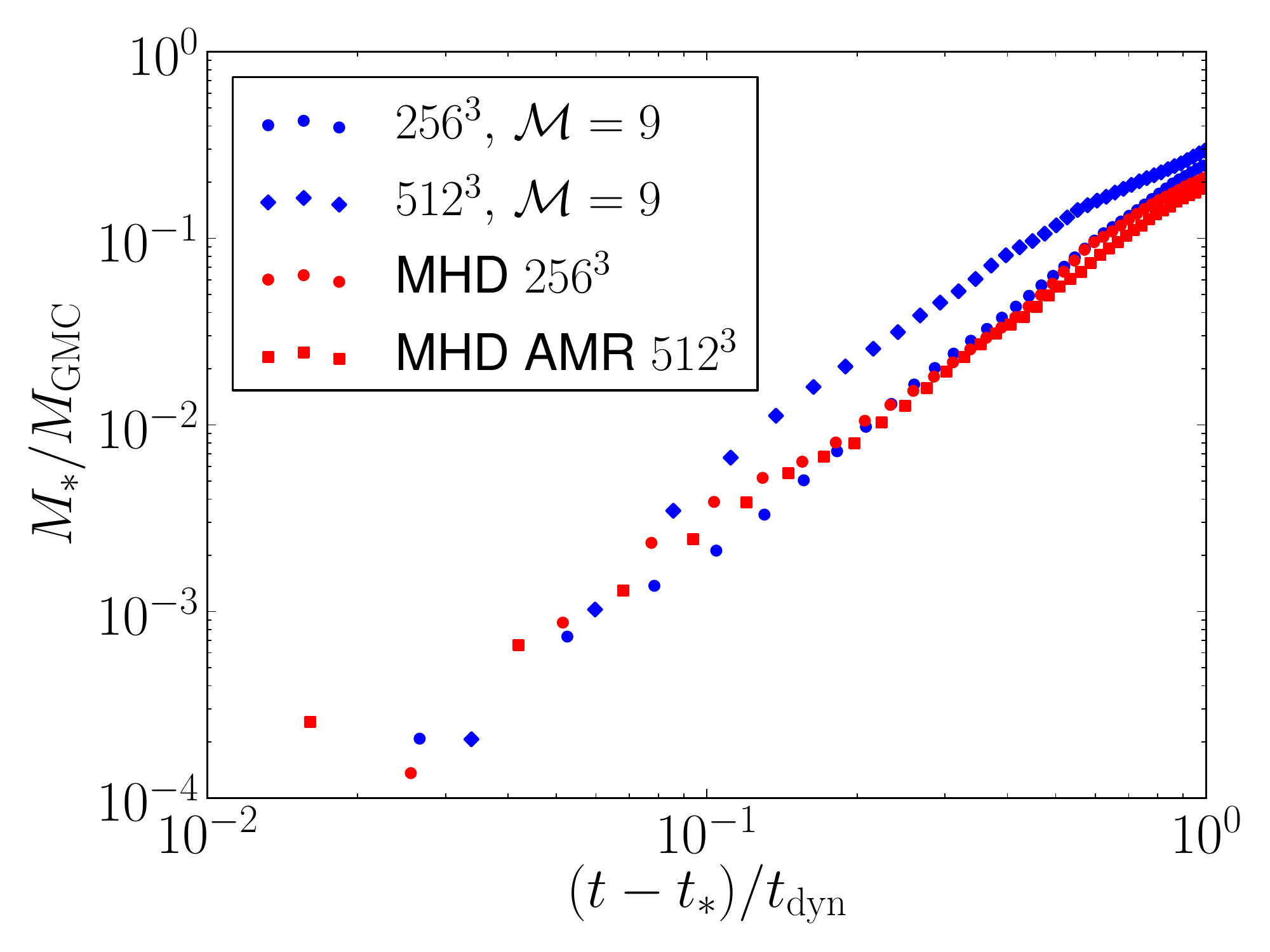}{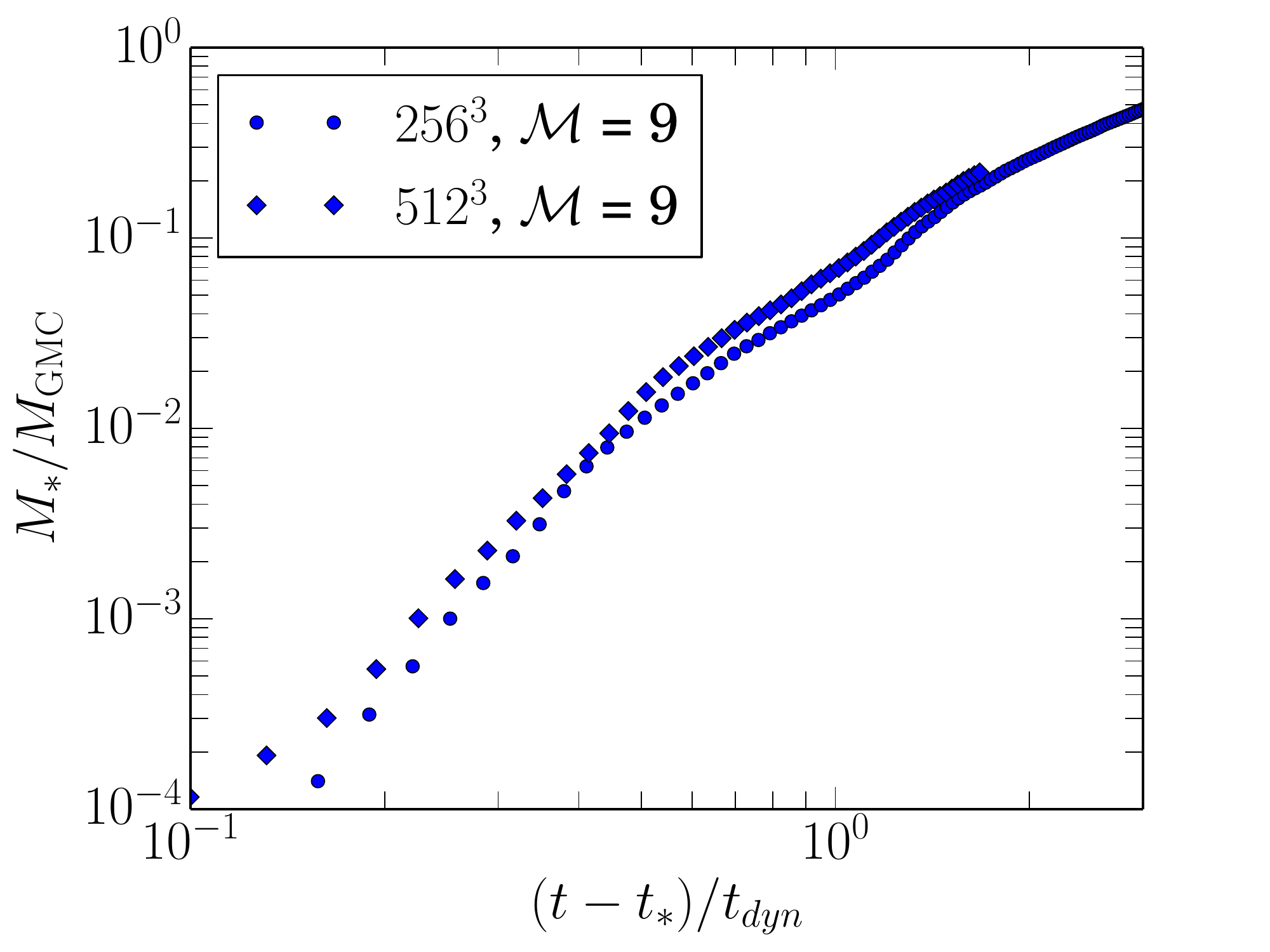}
 \caption{SFE as a function
   of $t-t_*$, where $t_*$ is the time when the first star particle
   formed.  Left: FLASH run. HD(blue points) and MHD (red points) are
   plotted at different resolutions: $256^3$ (circles), $512^3$
   (diamonds).
   Power-law fits ($M_*/M_{\rm GMC} = A (t-t_*)^{\alpha_p}$) to the
   HD runs give $\alpha_p =1.9$--2 for $0.003 < M_*/M_{\rm
   GMC} < 0.3$. In MHD runs, $M_*(t-t_*)$ grows approximately as $(t-t_*)^{2.5}$.
   Right: ENZO HD runs. 
   Fitting a power law in the range $0.015 < M_*/M_{\rm GMC} < 0.3$ 
   gives $\alpha_p = 2.31 \pm 0.02$ for $256^3$ and $2.22 \pm 0.01$ 
   for $512^3$. Even beyond $t_{\rm dyn}$, a fit to 
   $M_*/M_{\rm GMC} > 0.3$ gives $\alpha_p \simeq 1.6$. \label{fig:sfe}}
\end{figure*}

\subsection{Star Formation Prescription}
\label{ssec:sfp}

In our fiducial simulations, formation of stars (hereafter ``star
particles'') is governed by the Truelove criterion: a star is formed
at a grid point at which the Jeans length falls below four grid
cells~\citep{1997ApJ...489L.179T}.\footnote{The minimum number of
  cells to resolve Jeans length has been revised to $\sim$30 as four grid
  cells can overestimate SFE by $\sim$11\% \citep{federrath14_jeans}
  and underestimate the magnetic field intensity by at least an order
  of magnitude \citep{federrath11_jeans}.}  The Truelove criterion is
effectively a resolution-dependent density threshold criterion: stars
form if the cell has a density above
\bea
\frac{\rho}{\rho_{0}} &=& 740 \left(\frac{\rm{N}}{512}\right)^{2}
\left(\frac{16\rm{pc}}{\rm{L}}\right)^{2} 
\left(\frac{c_{s}}{265 \rm{m}~\rm{s}^{-1}}\right)^{2}\nonumber\\
&&\left(\frac{3\times 10^{-22} \g\cm^{-3}}{\rho_{0}}\right)
\label{eq:tlcrit}
\eea
where $\rho_{0}$ is the mean density. The material above this
density is collected onto a star particle, lowering the density in the
grid cell to the value given in Equation (\ref{eq:tlcrit}). Once 
stars form, they grow in mass via two channels: accreting gas within 
Bondi radius $GM_*/(c_s^2+v_T^2)$ and merging with other 
star particles when they are a half-cell distance away (two-cell distance for FLASH).

For comparison, \citet{2011ApJ...730...40P} form stars only in grid
cells where the density exceeds a fixed threshold of 8000$\rho_{0}$,
independent of the resolution. At $512^3$ resolution, this threshold
density is about 10 times that specified by the Truelove criterion
(Equation (\ref{eq:tlcrit})). However, we show in Section
\ref{ssec:sfe_evol} that the mass in star particles, $M_*(t)$, is
independent of the threshold criterion employed.

The default star particle methodology used in FLASH 4.0.1
is described in \citet{2010ApJ...713..269F}. However, we make some
significant modifications to reduce the computational load imposed by
the formation of hundreds to thousands of stars. As described in
\citet{2010ApJ...713..269F}, star particles are subcycled via a 
brute-force $O(N^2)$ algorithm, where $N$ is the number of star
particles. Computing the interactions between star particles
themselves is not generally time-intensive. However, there are
$O(N\times M)$ force evaluations between the gas and the star
particles and vice versa, where $M$ is the number of grid points.  In
addition, during each subcycle step, the particle's position must be updated across all the
operating cores.  We have found that handling gas-particle interactions is the slowest
step of the FLASH code.

To increase the efficiency of the FLASH code, we change the force
evaluation between the gas and the star particles and between the star
particles and the gas as follows.  First, we compute the force between
star particles using the same $O(N^2)$ algorithm. Second, we perform
the force evaluation on the gas by the star particles, by mapping the
particles to the gas grid and solving the Poisson equation. As the
Poisson solver is already executed to compute gas self-gravity, the
force evaluation between the star particles and the gas is done with
the minimal overhead of mapping the star particles onto the grid. 
Third, to solve for the force
on the star particles by the gas, we solve the Poisson equation
purely for the gas without mapping the star particles.

As pointed out by \citet{2010ApJ...713..269F}, this approach leads to
large errors in the dynamics of binary stars integrated over many
binary orbital periods.  However, we are interested here in the SFR; 
our tests show little difference between the SFRs obtained using our 
faster method as compared to
Federrath's method. We caution that the properties of the binary and
multiple stars formed, which we do not consider in this paper, will
likely be different using the two methods.

\section{RESULTS}
\label{sec:analy}

\subsection{Evolution of Star Formation Efficiency}
\label{ssec:sfe_evol}

After the turbulence has fully developed, we turn on the effects of 
self-gravity and star particle formation.  The global turbulent velocity
power spectrum remains unchanged, as was seen by
\citet{2012ApJ...750...13C} and \citet{2013ApJ...763...51F}.

Figure \ref{fig:sfe} shows how the total stellar mass in the
simulation volume, $M_*(t-t_*)$, evolves from the time $t_*$ at which
the first star particle forms. In both the pure hydro and MHD runs, we find the total
stellar mass is well described by a power law, $M_* \propto
(t-t_*)^{\alpha_p}$, with $\alpha_p\approx1.9$ in the range $0.003 <
M_*/M_{\rm GMC} < 0.3$ for FLASH (both HD and MHD) and $\alpha_p\approx2.2$ in the range
$0.015 < M_*/M_{\rm GMC} < 0.3$ for ENZO. $M_{\rm GMC}$ is the total
mass (both stellar and gas) inside the box. Including data points at lower $M_*/M_{\rm
  GMC}$ steepens the slope because the collapse dynamics is dominated
by gas rather than star particles; including data points at higher
$M_*/M_{\rm GMC}$ flattens the slope because stars are closely packed, interrupting the gas accretion. The exact lower and upper
limits of the fitting range are not well determined and chosen after
visual inspection to avoid obvious breaks in the power law.

This scaling is roughly the same over our rather limited range of different
resolutions. We note that $t_*$ depends on the star
formation criterion used and the strength of the initial magnetic
field, but we find that the power-law index does not depend strongly on $t_*$.

\begin{figure*}[!htb]
 \plottwo{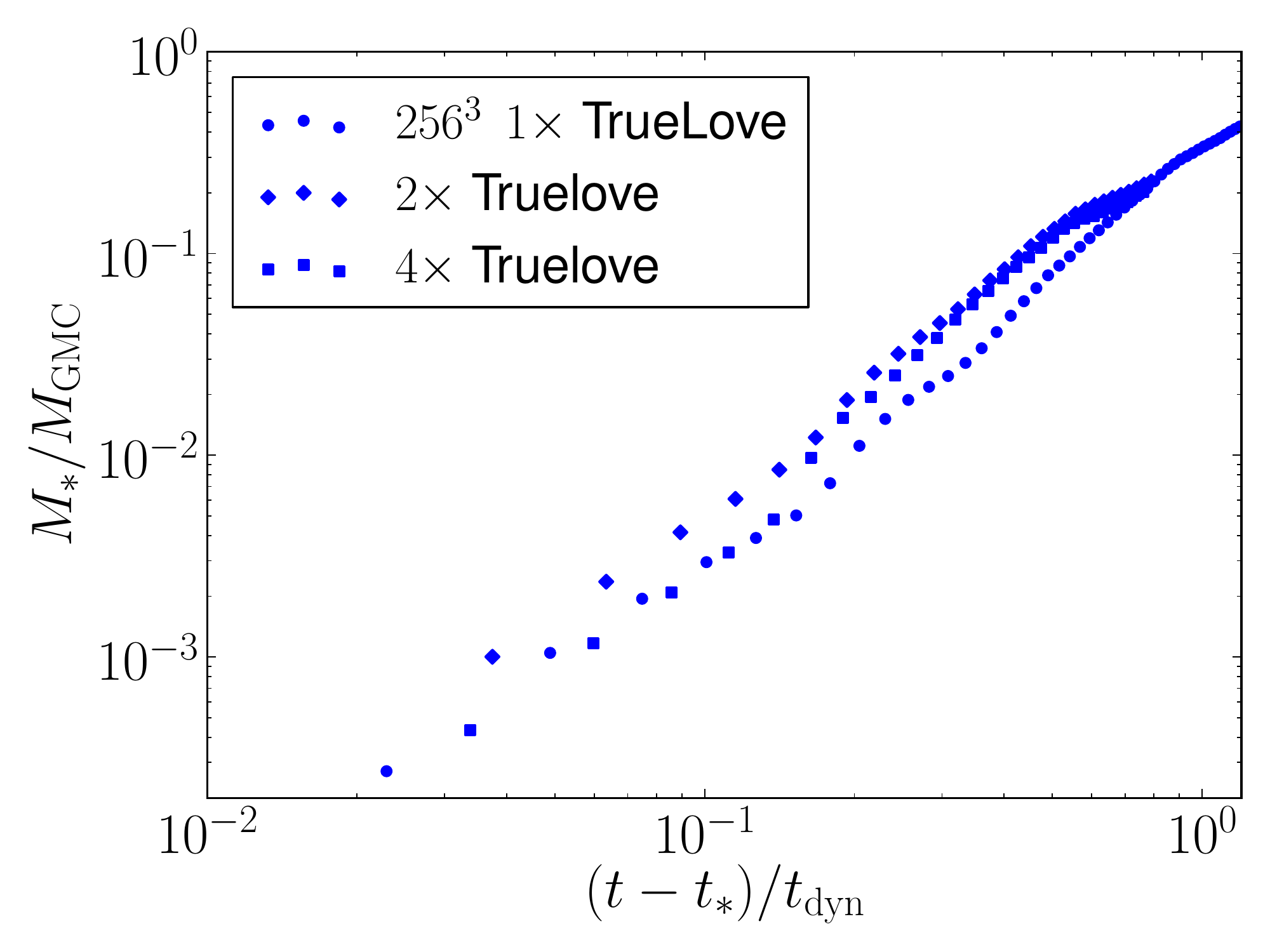}{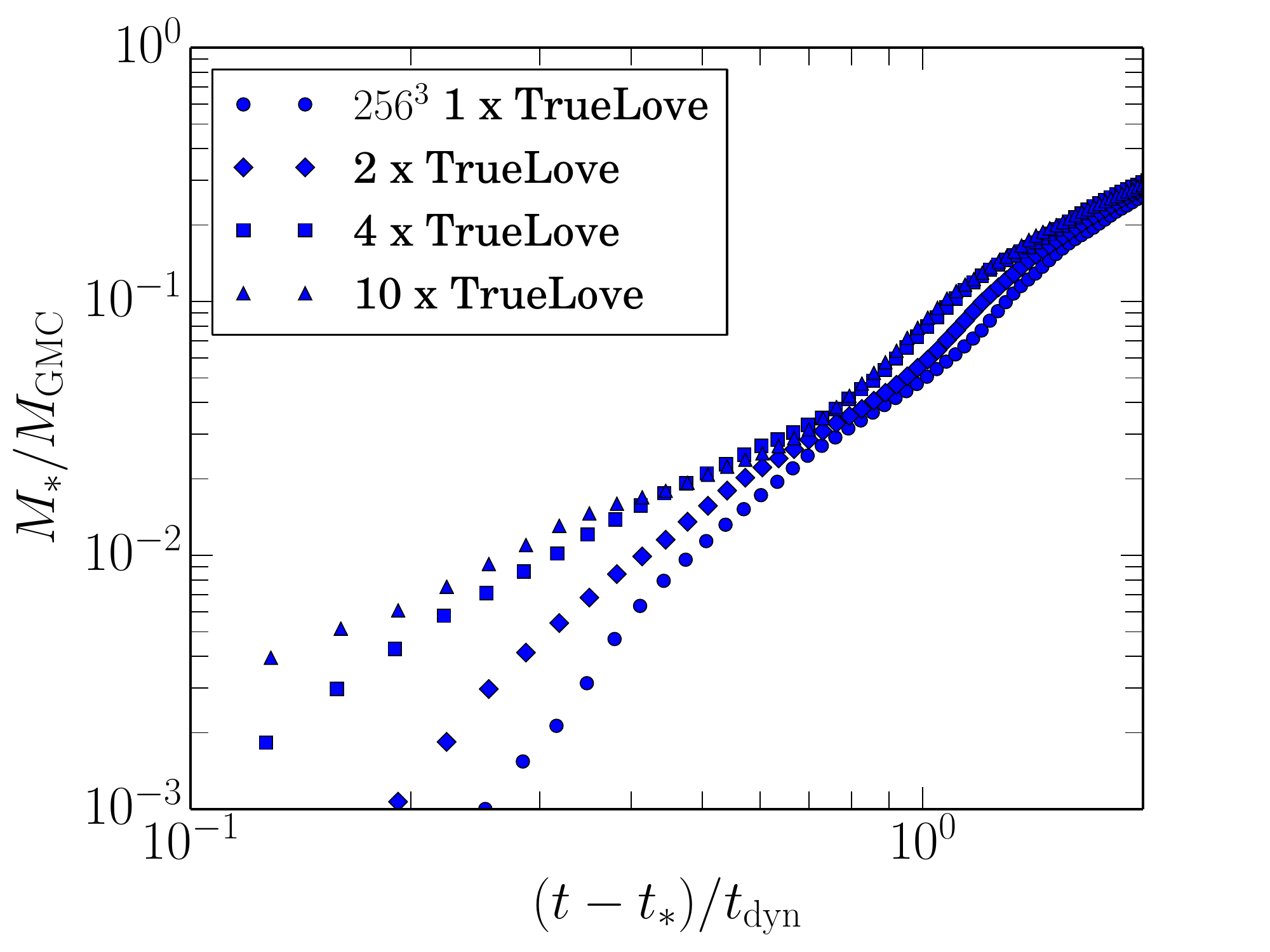}
 \caption{SFE as a function of $(t-t_*)$ at
   a resolution of $256^3$ for different star formation prescriptions.
   The left panel shows the result for a FLASH run while the right
   panel shows the result for an ENZO simulation.
   Plotted are results for our standard sink particle creation prescription 
	which uses the
   Truelove criterion (filled circles), and prescriptions that use
   2 times the Truelove density criterion (diamonds), 4 times the Truelove
   density (squares), and 10 times the Truelove density (triangles).
   Power-law fits ($M_*/M_{\rm GMC} = A
   (t-t_*)^{\alpha_p}$) to these points for $0.015 < M_*/M_{\rm GMC} <
   0.3$ show that $\alpha_p$ does not deviate significantly from 2 as a 
	result of varying the star formation prescription. Hence the SFR in 
	the simulation is controlled by the rate of
   collapse, not by our subgrid model of star
   formation. \label{fig:sfe_prescription}}
\end{figure*}

\begin{figure}[!htb]
        \plotone{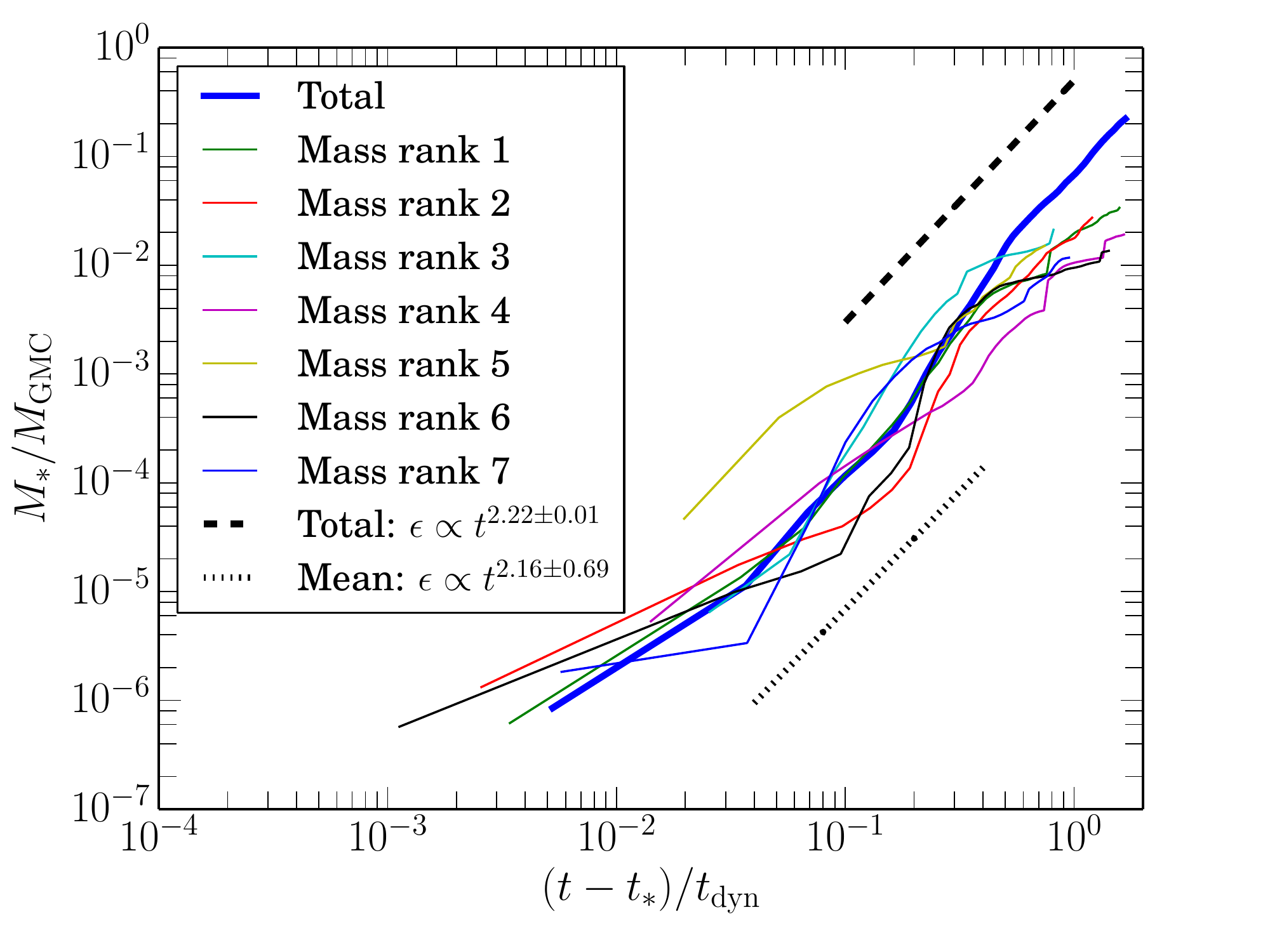}
	\caption{\label{fig:sfe_indiv_enzo}SFE vs. time
          for the seven most massive individual star particles in an ENZO run
          with a resolution of $512^3$. The characteristic time $t_*$
          in this plot is defined as the time at which the star
          particle plotted formed. The abrupt increases in stellar
          mass are produced by mergers of two or more star particles. The
          ``Total'' thick dashed line corresponds to the power-law fit
          to the SFE of the whole box, which in this run has an index
          of $\alpha_p=2.22\pm 0.01$. The ``Mean'' thick dotted line
          corresponds to the mean power-law fit to the SFE of the seven
          individual star particles. The power-law fit is performed in the range
          $5\times 10^{-5} \leq {\rm SFE} \leq 10^{-2}$, and results in
          $\alpha_p=2.16\pm 0.69$.}
\end{figure}
This result is robust to variations in the star
formation prescription. Figure \ref{fig:sfe_prescription} shows
$M_*(t-t_*)$ when we vary the density criterion for star particle
formation by scaling it relative to the Truelove criterion
(Equation (\ref{eq:tlcrit})).  We have used density thresholds that are 2, 4,
and 10 times the Truelove criterion; we find that $M_*\propto t^2$ in all
our tests. This explicitly demonstrates that the SFR
is independent of the subgrid model we use for the small-scale
physics of star formation.

Does $\alpha_p\approx2$ hold for an individual star particle as well?
We show the mass evolution of individual star particles in Figure
\ref{fig:sfe_indiv_enzo}. Not surprisingly, individual stars have a
more stochastic life history than the ensemble of stars, but it is
clear that individual stars also tend to grow as $M_*(t)\propto
(t-t_*)^{2}$. Here we track the seven most massive star particles for a
high-resolution run ($512^3$). Most of the sudden increases in the
masses of individual star particles are due to mergers with other star
particles.

\subsection{Density Probability Distribution Function}
\label{ssec:density_distr}
\begin{figure}
 \plotone{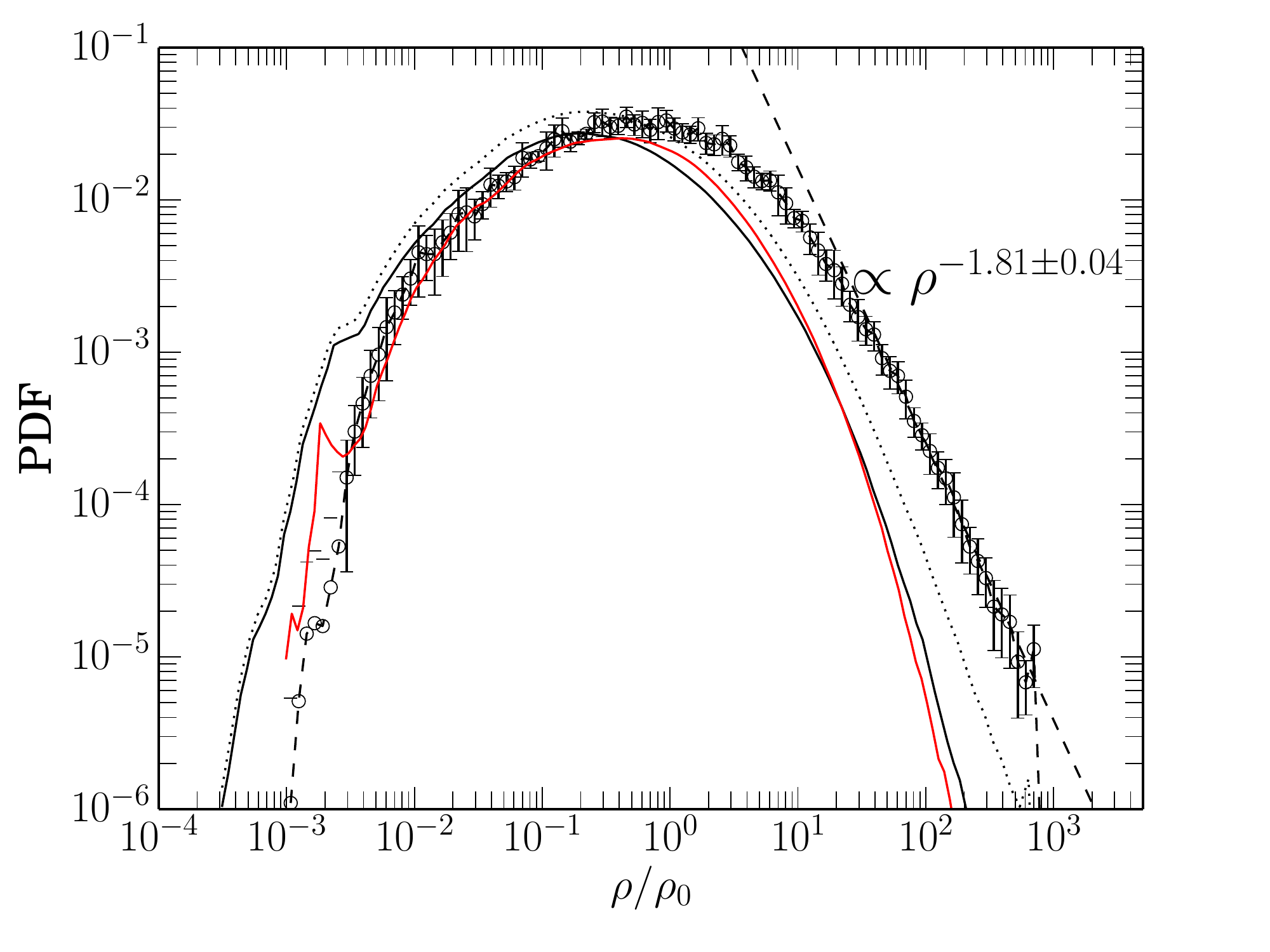}
 \caption{Volume-weighted density PDFs for SFE = 1\% for the entire 
   simulation box (dotted), excluding 3 pc radius spheres
   around star particles (solid), including only 3 pc radius spheres
   around star particles (dashed with data points), and 
   over the whole box before the gravity is turned on (solid red). 
   The excision of 3 pc radius spheres
   around star particles excludes the regions of highest density,
   resulting in a PDF that is more lognormal. Hence the power-law
   tail that we find in the entire volume is mainly associated with
   regions around star particles and is a result of gravitational
   collapse. The overlaid power law $\rho^{-1.81 \pm 0.04}$ is fitted
   for $30 \leq \rho \leq 300$. Note that the break at an overdensity
   of $\rho/\rho_0\sim700$ results from the star particle formation
   routine which removes gas from the grid and replaces it with star
   particles. This plot uses ENZO data.
\label{fig:rpdf_excs_enzo}} 
\end{figure}

As shown in Figure \ref{fig:rpdf_excs_enzo}, in gravity-free
supersonic turbulence, the volume-weighted density PDF is lognormal,
but when self-gravity is included, a power-law
tail emerges, corroborating the results of previous works \citep[e.g.,][]{klessen00,dib05,vazquez08,2008PhST..132a4025F,ballesteros11,cho11,2011ApJ...727L..20K,2012ApJ...750...13C,2013ApJ...763...51F}.  
While this has been noted by many
previous authors, only recently was it suggested by
\citet{2011ApJ...727L..20K} that this power law is due to regions that
collapse under self-gravity, and that the power-law exponent is
entirely determined by the density profiles of these collapsing
regions.

To check this suggestion, in Figure \ref{fig:rpdf_excs_enzo} we plot
the $\rm{PDF}(\rho)$ of regions undergoing gravitational collapse, of
regions largely unaffected by the gravity of the star particles, and
of the entire simulation box before and after the gas self-gravity is
turned on. We define the regions undergoing collapse as spheres each
having a radius of $3\pc$ centered on star particles. Gas lying outside
of these regions is considered noncollapsing. We see that the density
PDF of the noncollapsing regions matches well the PDF of the entire
box before the inclusion of gravity (i.e., the PDF is lognormal
without a power-law tail at high densities). This implies that regions
that do not undergo collapse retain the character of pure supersonic
turbulence. On the other hand, the density PDFs of collapsing regions
show a clear power law at high density. We find this power-law tail
$\rho^{-\beta}$ to scale as $\beta = 1.81 \pm 0.04$.

\begin{figure*}
\plotone{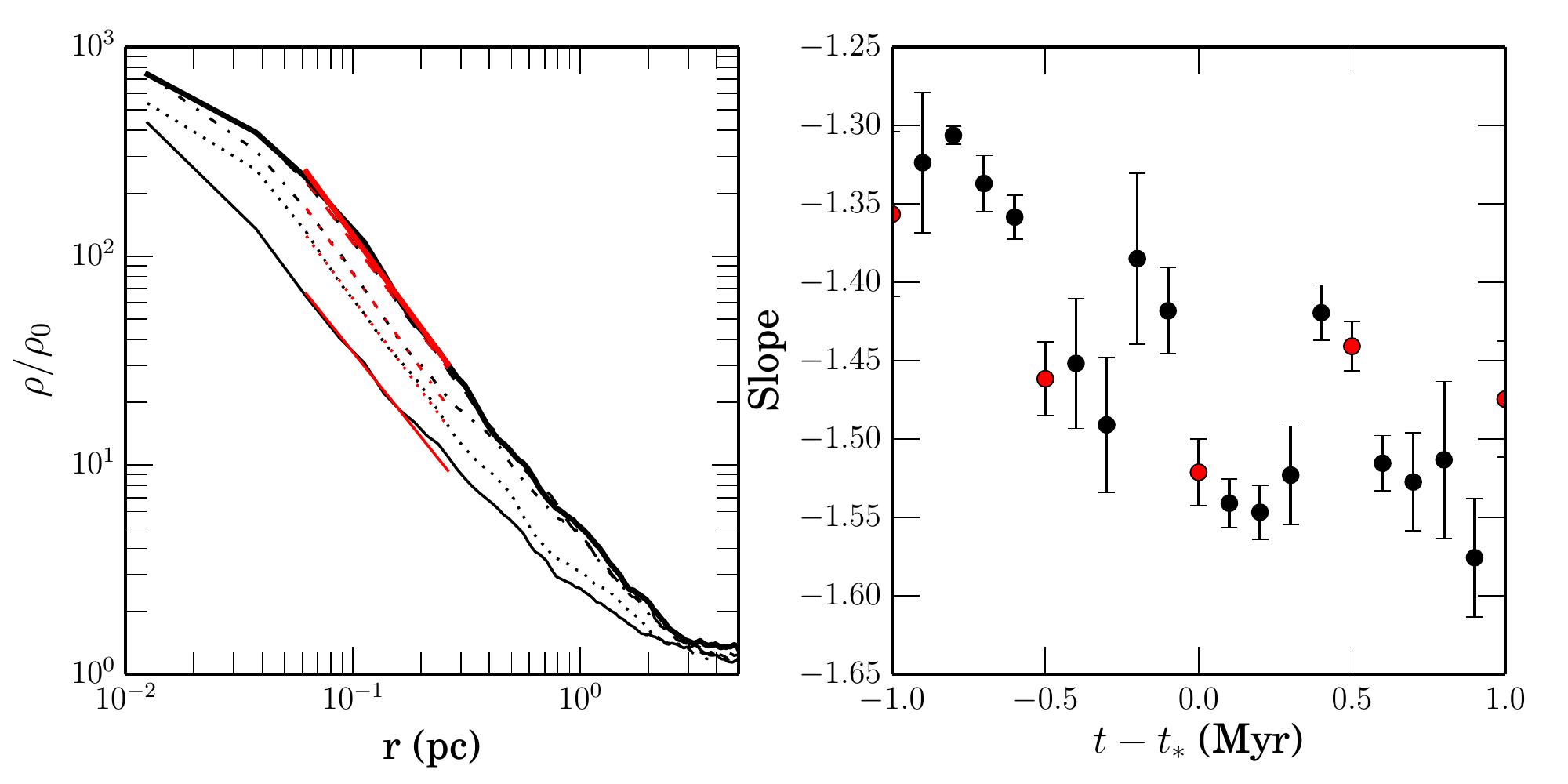}
\caption{\label{fig:star1_rhoprof} Left: radial density profiles
  around star particles. Plotted curves are averages over the 10 most
  massive star particles sampled at different times: 1 Myr before star
  formation (thin solid), 0.5 Myr before star formation (dotted), at
  the time of star formation (dot-dashed), 0.5 Myr after star
  formation (dashed), and 1 Myr after star formation (thick
  solid). Red lines are power-law fits to the profiles in the range
  $30 < \rho/\rho_0 < 300$. Right: fitted power-law indices for the
  density profile, averaged over the 10 most massive star particles, in the
  range $30 < \rho/\rho_0 < 300$ as a function of time. Red circles
  illustrate the fits to the profiles shown in the left panel. This
  plot uses ENZO data.}

\end{figure*}

If the density is spherically symmetric and follows a power law $\rho
\propto r^{-k_{\rho}}$, then $\beta = 3/k_{\rho}$. Figure
\ref{fig:star1_rhoprof} shows the evolution of the power-law index of
the density profile averaged over 10 massive star particles:
$k_{\rho}$ starts at $\sim1.30$ before the formation of star particles
and then increases to $\sim1.55$ after star formation. The expected
range in $\beta$ given the final $1.45 \lesssim k_{\rho} \lesssim
1.62$ is 2.07--1.85 which agrees with the numerically determined
$\beta = 1.81$. The small discrepancy between the expected and
observed $\beta$ values is likely due to the combined effect of star
particles at different stages of their formation and
evolution. Furthermore, a simple power law is only a rough
approximation to the run of density; as illustrated in Figure
\ref{fig:star1_rhoprof}, density profiles are not perfect power
laws. While our inferred $\beta$ is steeper than the values
\citet{2011ApJ...727L..20K} report, this is likely because they have
higher resolution and they use a different fitting range.

The fact that the density PDF in a self-gravitating gas is lognormal
if and only if local density peaks are excised, combined with the very
prominent power-law tails seen in the PDF calculated in small volumes
around those same peaks and that these slopes of power-law tails are
consistent with the slope of the averaged density profile, confirms
the suggestion of \citet{2011ApJ...727L..20K} that the power-law tail is due to
gravitationally induced collapse.

\subsection{Importance of Gas Self-gravity}
\label{ssec:evolving_rhoPDF}

Simulations of star formation must account for several kinds of
gravitational interactions: the self-gravity of gas on gas, the
self-gravity of stars on stars, and the gravity between gas and
stars. Star particle creation routines provide a fourth model for
gravity on sub-grid scales. Our sub-grid model is ideally suited to
testing the idea that star formation results from collapse of gas in a
lognormal density PDF above a density on the scale of the sonic
length, as long as the sonic length is
resolved.

In this subsection, we systematically turn on and off each of these
four kinds of gravitational interactions to determine their relative
impact on the time evolution of SFE and the
density PDF. 

\begin{figure}
        \plotone{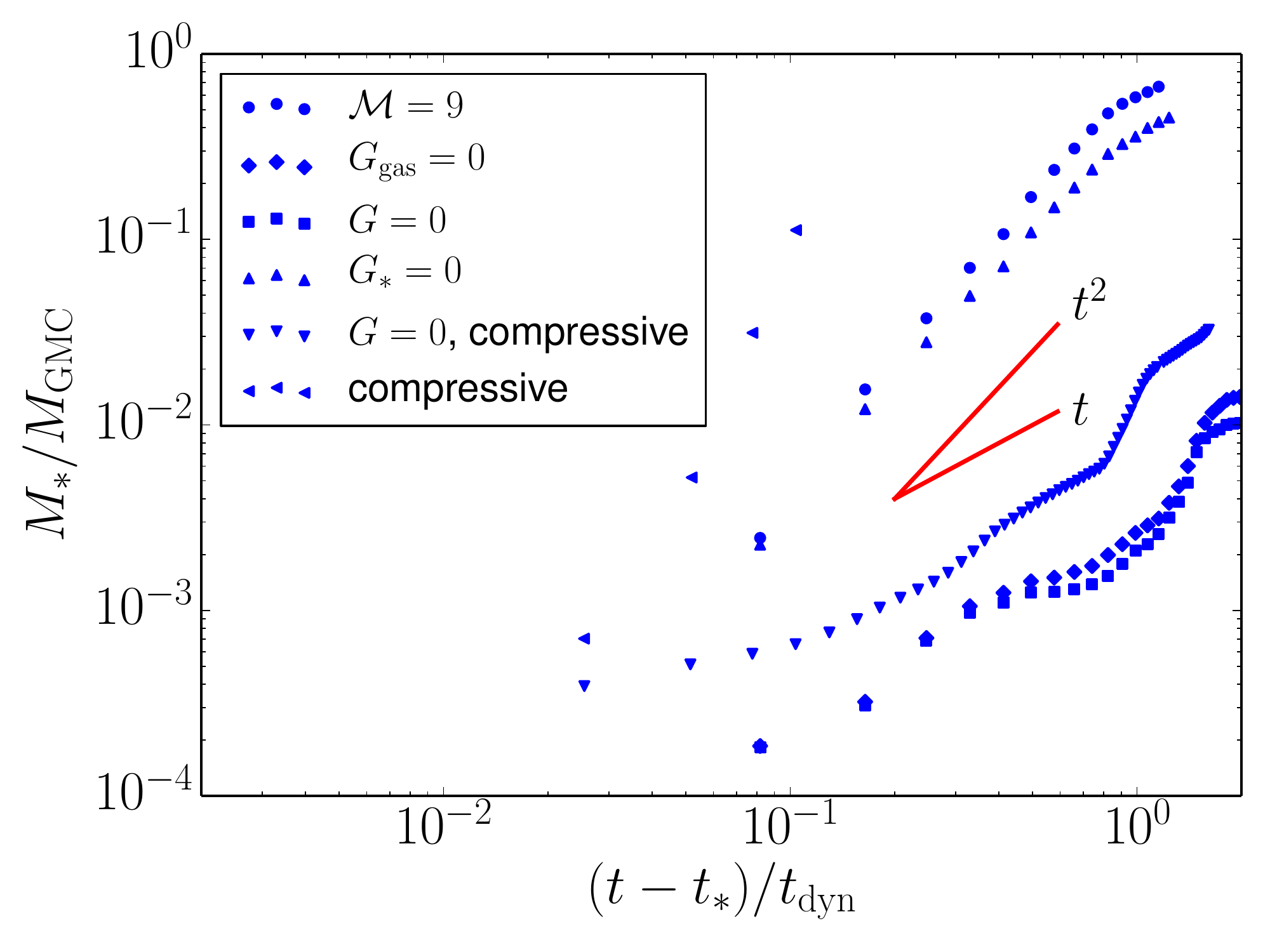}
	\caption{\label{fig: no self-gravity}SFE vs. time in six
          runs with a resolution of $256^3$. The circles depict runs
          in which all gravitational interactions are accounted
          for. The diamonds exclude gas self-gravity. The squares and
          left-pointing triangles depict runs where all gravitational
          interactions except for that inherent in the star formation
          routine are excluded. The up or down triangle runs exclude
          stellar gravity. The driving is purely solenoidal
            except in the run depicted by the left-pointing triangles,
            and in the run denoted by inverted triangles, both of
            which have purely compressive driving, with the latter
            excluding gas self-gravity.  The runs that account for
          the self-gravity of the gas have $M_*(t)\sim t^{\alpha_p}$
          with $\alpha_p\approx2$ for solenoidal forcing, and
          $\alpha\gtrsim3$ for compressive forcing. Linear fits to these runs at late
          times find $\eff\approx 0.3$--0.5.  In contrast, runs that do
          not include self-gravity have $\alpha_p\approx1$, yielding a
          constant $\eff\approx 0.003$ or slightly higher for
          compressive forcing. This plot uses FLASH data.}
\end{figure}

Figure \ref{fig: no self-gravity} demonstrates that neglecting
  the gas self-gravity leads to a much slower SFR,
  roughly by a factor of 10, and a linear mass growth rate
  with $\alpha_p\approx 1$. In contrast, the gravity due to stars does
  not have such a dramatic effect on the star formation: the rate
  decreases by less than a factor of two, with an unchanged power-law
  exponent $\alpha_p\approx2$. Simulations employing
  compressive turbulence driving tend to have SFE about factors of 10
  larger than simulations under solenoidal turbulence
  \citep{2012ApJ...761..156F}, a result we reproduce here 
	(left-pointing triangles in the figure).  However, Figure \ref{fig: no
    self-gravity} shows that even under purely compressive turbulence
  driving, in the absence of gas self-gravity, SFE remains about
  a factor of 10 lower than simulations with gas self-gravity under
  purely solenoidal turbulence.

\begin{figure}
\plotone{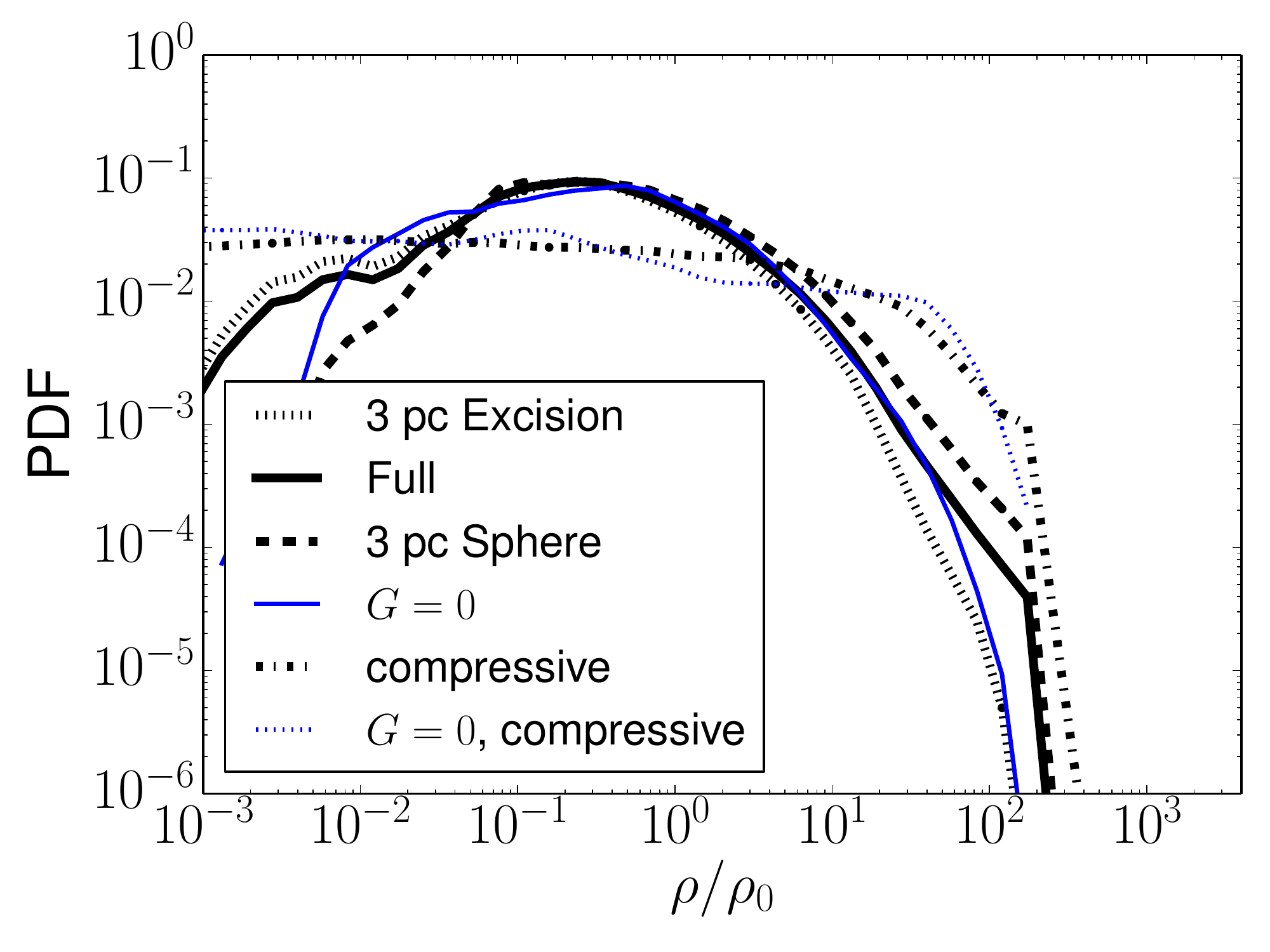}
\caption{\label{fig:PDF_exclude_no_grav} Volume-weighted density
    PDFs for SFE = 1\% for different turbulent driving (solenoidal
    v.s.~compressive) and for different volumes. First, the results
    of a solenoidally driven run including all types of gravitational
    interactions for (a) the entire simulation box (thick black solid
    line), (b) the volume contained in 3 pc spheres around star
    particles (thick black dashed line), and (c) the entire box
    excluding 3 pc spheres around star particles (black dotted
    line). The thin solid blue line shows the result of a run with no
    gas self-gravity. The remaining two lines show results for
    compressively driven turbulence, resulting in much broader density
    PDFs: the dotted blue line depicts a run with no self-gravity,
    corresponding to the downward-pointing triangles in Figure
    \ref{fig: no self-gravity}, while the dot-dashed line depicts a
    run including gas self-gravity (corresponding to the left-pointing
    triangles in Figure \ref{fig: no self-gravity}) The fact that
    there is only a very slight difference between the PDFs of the two
    compressively driven runs is likely a result of the limited
    spatial resolution; the turbulent cascade does not have enough
    spatial range to relax to a mix of compressive and solenoidal
    turbulence. Despite the similarity in the high-density tail of the
two PDFs, the large difference in the SFR between the
two runs shows that the velocity around star-forming regions is much
larger in the run including self-gravity.}
\end{figure}

Figure \ref{fig:PDF_exclude_no_grav} shows that the power-law density
tail does not appear in the absence of gas self-gravity in simulations
with solenoidal driving, consistent with our contention that the tail
is due to gravitationally induced collapse driven by the gas-on-gas
potential rather than the star-on-gas potential.

Note, however, that the figure also shows that compressive
  turbulent driving broadens the lognormal density distribution. The
  density PDFs of the two compressively driven runs are depicted by
  the dotted and dot-dashed lines. This broadening occurs whether the
  gas self-gravity is included or not; because of our limited
  resolution, it is difficult to determine whether there is a
  power-law tail to either of these two PDFs. This broadening is 
	reminiscent of that seen in observations of molecular gas compressed by
  ionization feedback from HII regions \citep{tremblin14}. Numerically, 
	the sensitivity of density distribution on the mode of turbulence driving 
	has been extensively studied and discussed by \citet{federrath08} and 
	\citet{federrath10aa}.

While the difference between the PDFs of the two compressively
  driven simulations is difficult to see in Figure
  \ref{fig:PDF_exclude_no_grav}, the rapid rate of star formation in
  the gravity-on case (left-pointing triangles in Figure \ref{fig: no
    self-gravity}), compared to the gravity-off case, (downward-pointing
		triangles in that figure) shows that the rate of mass flow from
  large scales to small scales is far faster in the gravity-on
  case. We interpret this to mean that the much broader density PDF in
  compressively driven turbulence is not {\em directly} responsible for the more
  rapid star formation rate seen in previous simulations of star
  formation in previous work. Instead, it is the
  increased infall velocity, produced by the initially denser
  post-shock gas's self-gravity, that
  leads to the high SFR, and not the compressive
  driving by itself. We suspect that our low resolution might play a
  role in the very strongly enhanced SFR we find in
  the gravity-on compressively driven case; were we able to follow the
flow to yet smaller scales, the enhanced turbulence in the
compressively driven case might be seen to slow the infall velocity on
those smaller scales, compared to a noncompressively driven case. We
are investigating this point further.

These experiments show that gas self-gravity is the primary driver of
the rapid SFR seen in our simulations, and of the
power-law tails in density PDFs. Models of star formation must,
therefore, include not only the small-scale potential produced by
stars, but also the large-scale ($\gtrsim 0.1\pc$) potential produced
by gas. Models that neglect the large scale effects of self-gravity,
e.g., those that assume a log-normal density PDF and no feedback
effects, will underestimate the SFR by factors of 10 or more.

\subsection{Velocity Profiles}
\label{sec:vprof}

\begin{figure}
 \plotone{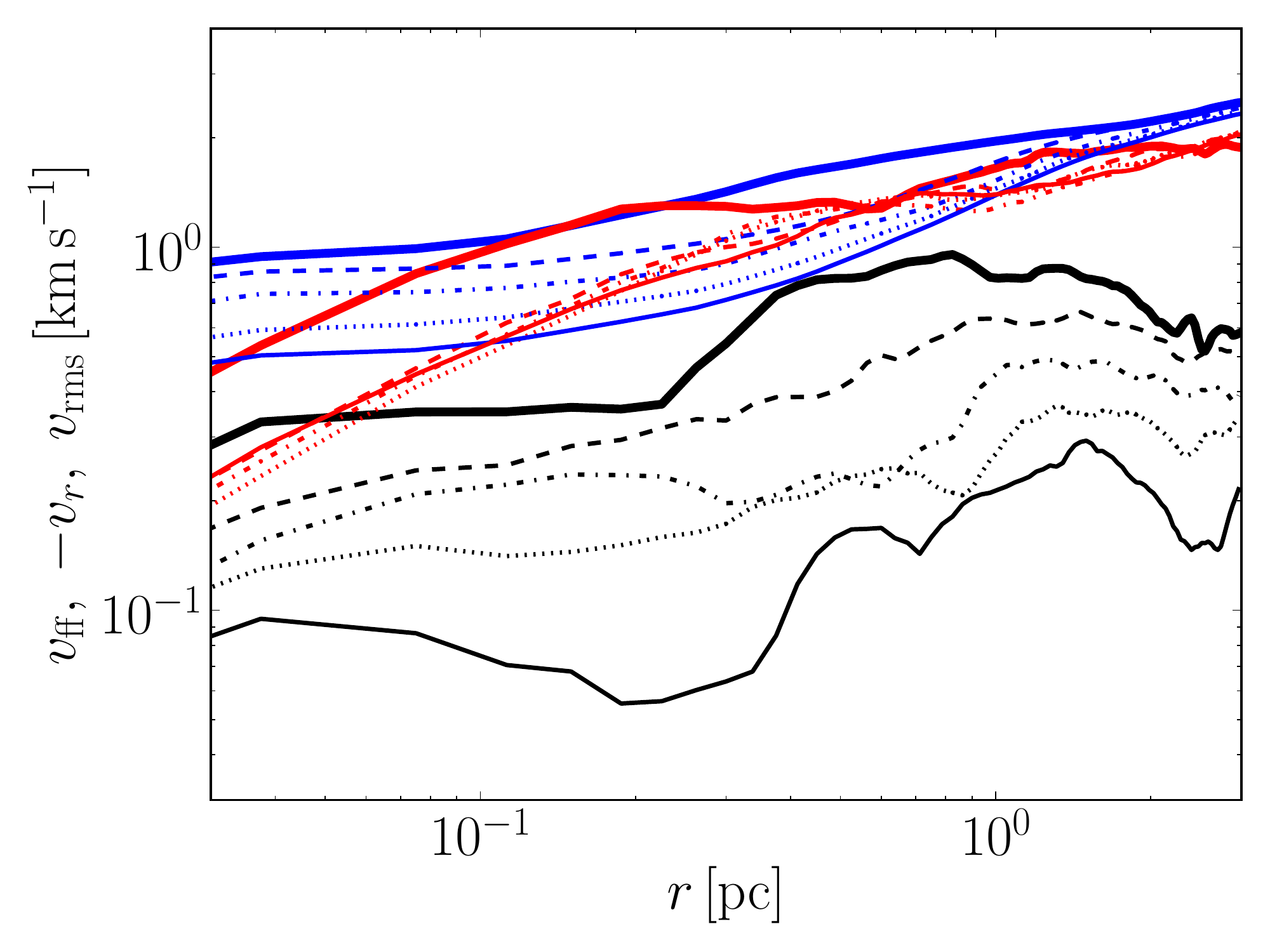}
 \caption{\label{fig:velocities_flash} Various velocities plotted
   vs.~radius from a local density peak, averaged over 10
   peaks. The black curves show $v_r$, the infall velocity onto the
   density peak, while the red curves show the rms velocity. Power-law
   fits to $v_{\rm rms}(r)\sim r^p$ yield $p=$0.25--0.35, significantly
   smaller than the exponent $p=0.5$ measured in the bulk of the
   box. At late times, $p\approx0.1$ near density peaks in some
   runs. The blue curves show $v_{\rm ff}$. The line styles
   correspond to different times, measured from the time of formation
   of the first star in each density peak: thin solid, dotted,
   dot-dash, dash, and thick solid lines correspond to $-0.5$, $-0.3$, $-0.1$,
   $0.2$ and $0.6 \Myr$ before (negative) or after (positive) the first star forms. 
   The bottom panel shows that $v_r < v_{\rm rms} < v_{\rm ff}$ at all radii, suggesting 
    that a substantial fraction of the gas near the local density peak is bound out to at least $\sim$3pc.}
\end{figure}

We now turn to the velocity profiles around these collapsing
regions. In Figure \ref{fig:velocities_flash}, we plot the
mass-averaged infall velocity $v_r$, the free-fall velocity $v_{\rm
  ff}=\sqrt{2GM(<r)/r}$, and the rms velocity 

\be 
v_{\rm rms}({\bf r})\equiv \sqrt{\left\langle
({\bf v}-{\bf v_r})^2({\bf R}+{\bf r})-{\bf v}^2({\bf R})
\right\rangle}.
\ee 

\noindent Here ${\bf r}$ measures displacement relative to a reference 
position ${\bf R}$.

Around density peaks, we find that $v_{\rm rms}\propto r^{0.3}$, flatter than the
background turbulent velocity $v_{\rm T}\propto r^{0.5}$. We attribute the
enhanced turbulent velocity and flatter slope in the vicinity of
density peaks to the conversion of gravitational potential energy into
turbulent energy. The kinetic energy of the bulk inflow of gas also
contributes to the driving of turbulence, but as Figure
\ref{fig:velocities_flash} shows, its effect is minimal compared to
gravity. The idea that gravitational collapse can
drive turbulence has a long history
\citep[e.g.,][]{1953ApJ...118..513H,1982ApJ...258L..29S}, and it has been seen
in recent numerical studies
\citep[e.g.,][]{2010ApJ...721L.134S,2011ApJ...731...62F}. However, in the
latter two papers, the fraction of energy going into turbulent motions
is small compared to that going into radial infall. In our case, the
energy in turbulent motion is substantially larger than that in the
radial infall. The difference is likely in the initial conditions: 
\citet{2010ApJ...721L.134S} and \citet{2011ApJ...731...62F} start with a 
smooth, spherically symmetric density distribution with transonic
turbulence.

\begin{figure}
 \plotone{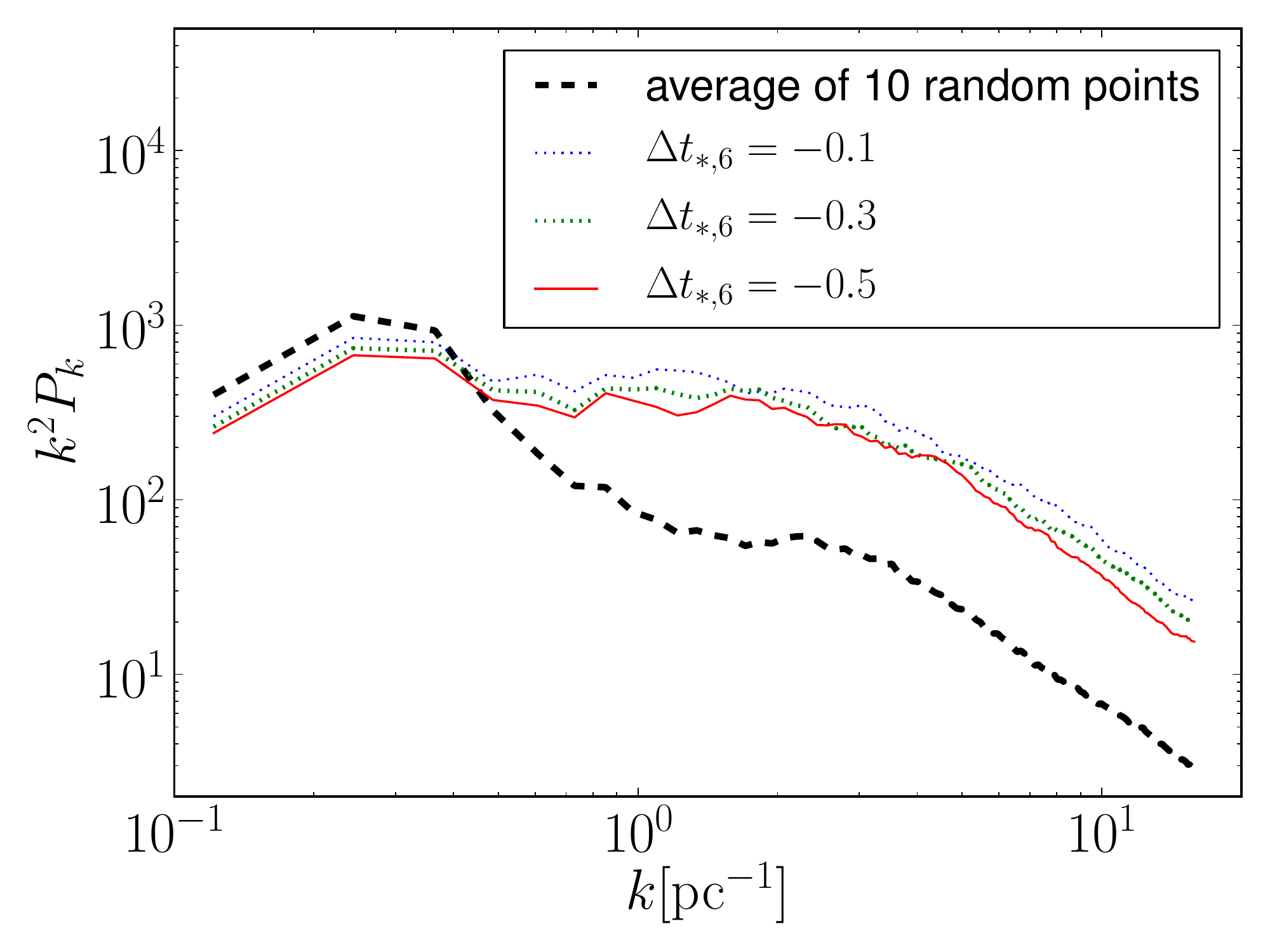}
 \caption{\label{fig:turb_around_star}Compensated velocity power
   spectrum inside an $(8\pc)^3$ cube around a local density peak,
   calculated after applying a Gaussian
   window function with $2\pc$ variance. The plotted 
   curves represent an average over three density peaks sampled at times
   0.1 Myr (blue dashed), 0.3 Myr (green dashed), and 0.5 Myr (red
   solid) before the
   first star forms in each peak. The solid black curve shows the
   power spectrum around five random points in the simulation box,
   calculated with the same window function. There is more
   power on small scales in regions around local density peaks than in
   the box as a whole.}

\end{figure}

Another way to see that the turbulent velocity is
enhanced only near local density peaks is to study the velocity power
spectrum around these points. Figure \ref{fig:turb_around_star} shows
the power spectrum calculated in cubes $8\pc$ on a side, centered
around local density maxima and also around random points away from
these maxima. Each power spectrum is calculated after  applying a
Gaussian window function with $2\pc$ variance centered on
the point of interest. The velocity power spectra around density
maxima have much more power on small scales than either the global
power spectrum (Figure \ref{fig:turb_b4grav}) or the power spectra
calculated in an identical manner around random points in the box.

\subsection{Collapse Geometry}
\label{sec:geometry}
\begin{figure*}
\epsscale{0.8}
\plotone{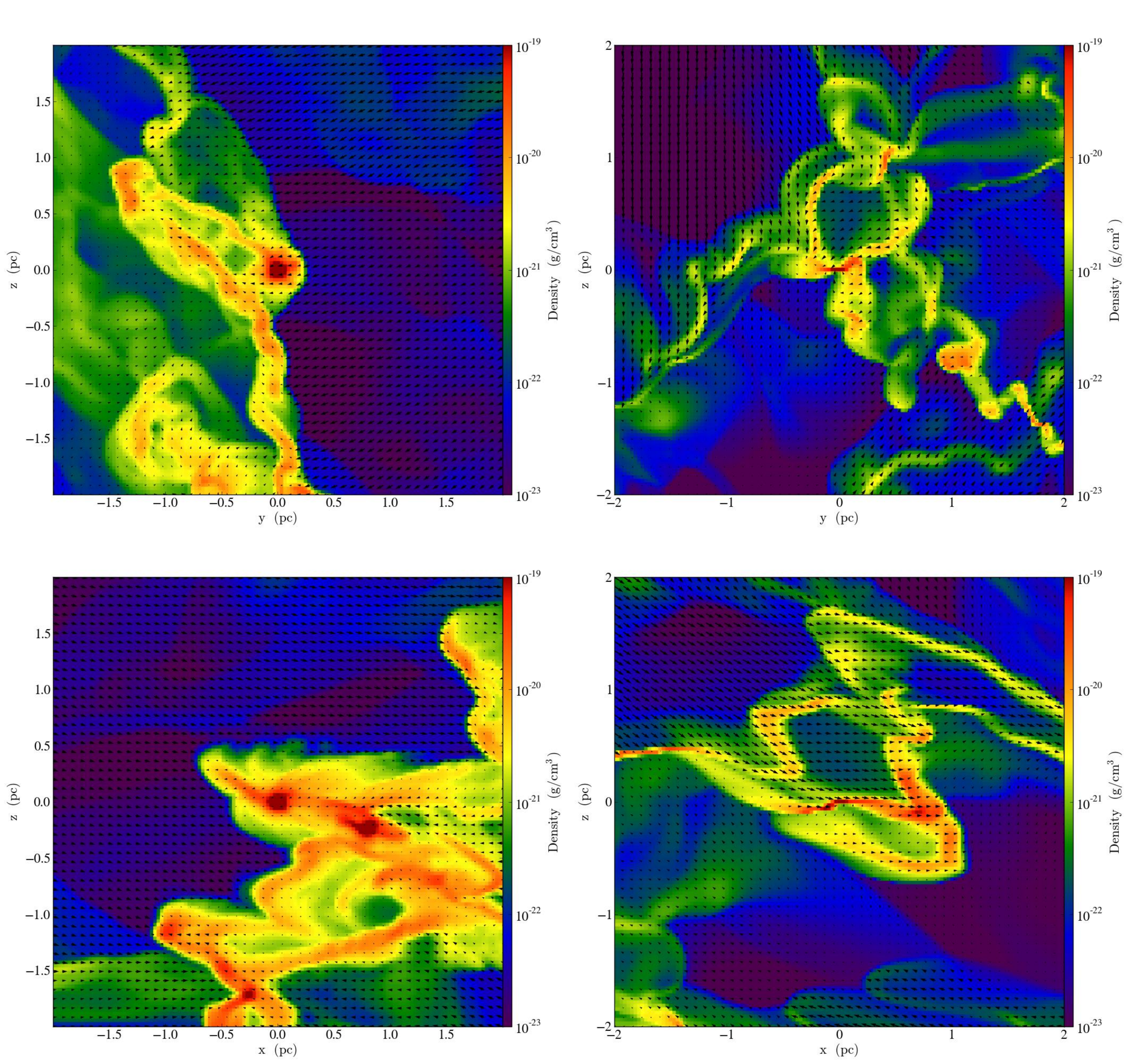}
\caption{\label{fig:density picture} Density (shown by colors) and
  velocity (shown by arrows, with the size proportional to the
  magnitude of the projected velocity) in the vicinity of a
  local density peak, 0.1 Myr before a star forms. The left two panels
  are for a FLASH run, while the right two panels are for an ENZO run. The upper
  panels shows slices one cell thick ($\sim0.03\pc$) in the
  $yz$-plane, and the lower panels show similar slices in the
  $xz$-plane, both centered on
  the location where the star will form. We find stars to form in a filament
  (orange) or in regions where multiple filaments converge. The apparent 
  difference between FLASH and ENZO density structure is due to 
  the more dissipative solver used by ENZO, as well as the different turbulence 
  driving scale. The online journal contains accompanying video files 
	(Figure 11a for FLASH runs and Figure 11b for ENZO runs) that show the density 
	projection of the entire simulation box evolving from the initial turbulence driving 
	to the end of the simulation after the formation of star particles.}
\end{figure*}

Figure \ref{fig:density picture} shows density slices in $xz$- and
$yz$-planes shortly before the first star forms. The figure
illustrates the well-known result that simulated stars form in
filaments or sheets, which in this case we define by gas with
$\rho>3\times10^{-20}\g\cm^{-3}$ (colored orange), equivalent to
  number density $n\sim 10^4\cm^{-3}$, the density of the interstellar medium
  filaments observed by \citet{peretto14}. The arrows show the projected velocity
of the gas, and illustrate the convergent nature of the flow in the
vicinity of the density peaks. The filaments in our simulation
  have length scale $\sim$1--2 pc, width $\sim$ 0.1--0.3 pc, total mass
  $\sim$100--200 $M_\odot$ (not shown), and velocity dispersion
	$\sim$1--2$\kms$ (see Figure \ref{fig:velocities_flash}), similar to
  the properties of the SDC13 infrared dark cloud \citep[see][their
    Figures 1, 2, 4 and Table 3]{peretto14}, and to the properties of the 
		IC 5146 {\it Herschel} filaments \citep{arzoumanian11}. Figure
  \ref{fig:velocities_flash} shows that the gas inside these high-density
  structures is by and large bound: the local turbulent velocity
  around density peaks is at or below the free-fall velocity at
  all radii at all times, while the bulk velocity is well below the
  free-fall velocity.

\epsscale{1.0}

\begin{figure}
\plotone{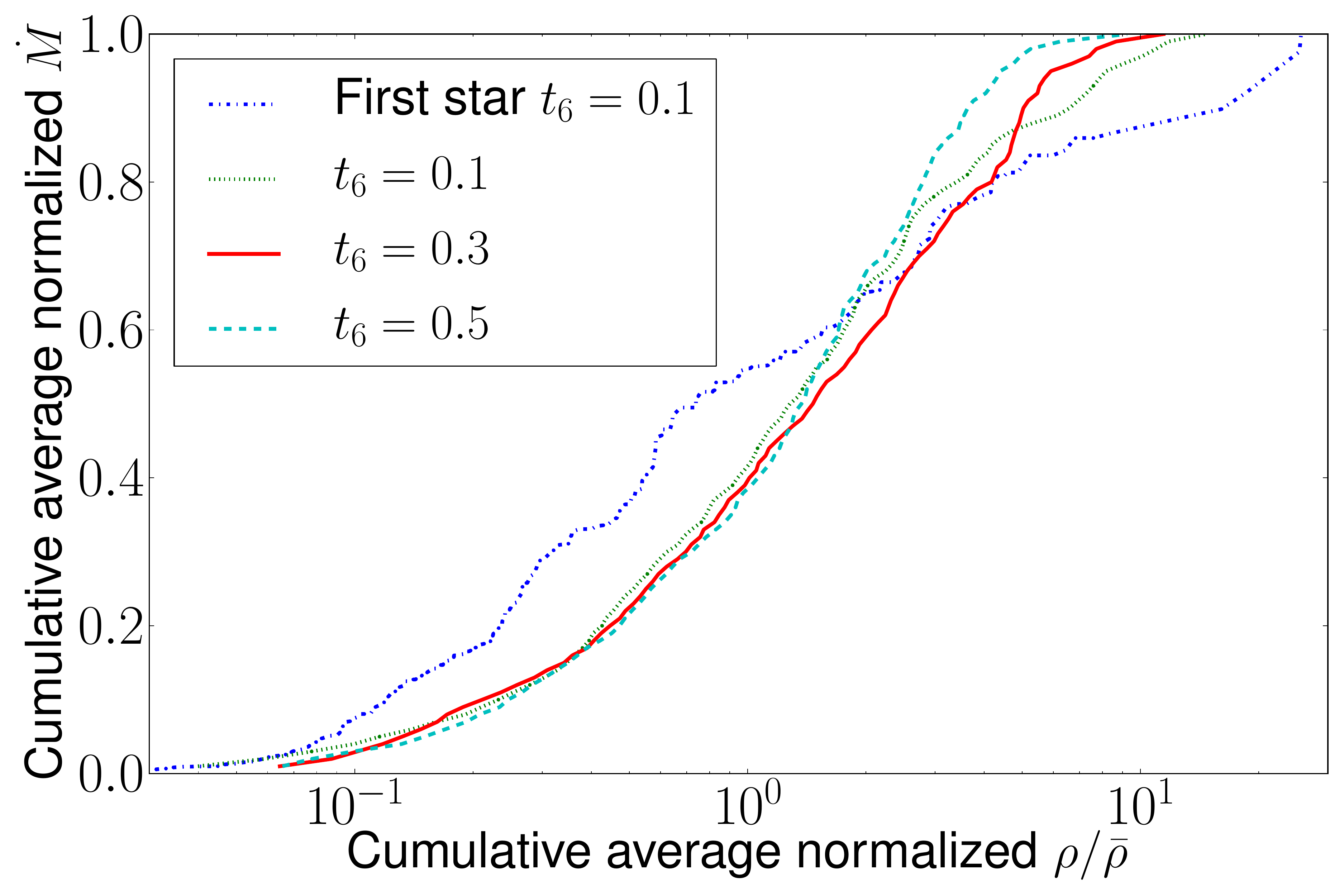}
\caption{\label{fig: cum_mdot_vs_density_normalized} Cumulative
  mass accretion rate, averaged over 10 star particles in a FLASH
  run. The rates are calculated $10^5$ yr (dotted blue line),
  $3\times10^5$ yr (solid green line), and $5\times10^5\yr$ (dashed
  red line) after the relevant star particle forms. The blue
  dot-dashed line depicts the cumulative mass accretion rate onto the
  first nascent star. The accretion rate is plotted as a function of
  $\rho/\bar\rho$, where $\bar\rho=3\times10^{-21}\g\cm^{-3}$ is the
  average density inside the spherical shell at $0.5\pc$. Roughly half
  the accretion rate is due to gas with a density below the mean
  density of the shell, indicating that accretion is not dominated by
  accretion through filaments; we define a filament by
  $\rho\ge3\times10^{-20}\g\cm^{-3}$, i.e., $n\approx10^4\cm^{-3}$,
  comparable to the density $n\approx3\times10^4\cm^{-3}$ in the
  filaments discussed by \citet{peretto14}. }
\end{figure}

Figure \ref{fig: cum_mdot_vs_density_normalized} quantifies the source
of mass falling onto stars. We plot the mass accretion rate $dM/dt$
across a spherical shell with radius $0.5\pc$ centered on the local
density maximum for 10 star particles at different times. In this
plot, we calculate $dM/dt$ for each cell in the shell and order them
by the density in that cell. We then plot the cumulative $dM/dt$ as a
function of density divided by the mean density in the shell,
$\bar\rho\approx3\times10^{-21}\g\cm^{-3}$
($n\approx10^3\cm^{-3}$). The figure shows that half the accretion
comes from regions whose gas densities are around or below the mean
shell density, whereas only a small fraction of material comes from
high-density regions, i.e., filaments. 

As a further illustration of this point, we also plot the mass
accretion rate for the first star particle, which is shown in the left
panels of Figure \ref{fig:density picture}. Once again, the volume
average density in the shell is
$\bar\rho(0.5\pc)=3\times10^{-21}\g\cm^{-3}$, while the density in the
filament is $\rho_{\rm filament}\approx3\times10^{-20}\g\cm^{-3}$, as
can be seen in Figure \ref{fig:density picture}. Only a small fraction
of accretion ($\lesssim 20\%$) proceeds through high-density regions,
which we are identifying with the filaments seen by
\citet{peretto14}. In fact, 80\%--90\% of the total mass inside
 a 2 pc radius sphere centered on the local density maximum resides in
  structures with $\rho \leq \rho_{\rm filament}$.

\begin{figure}
\plotone{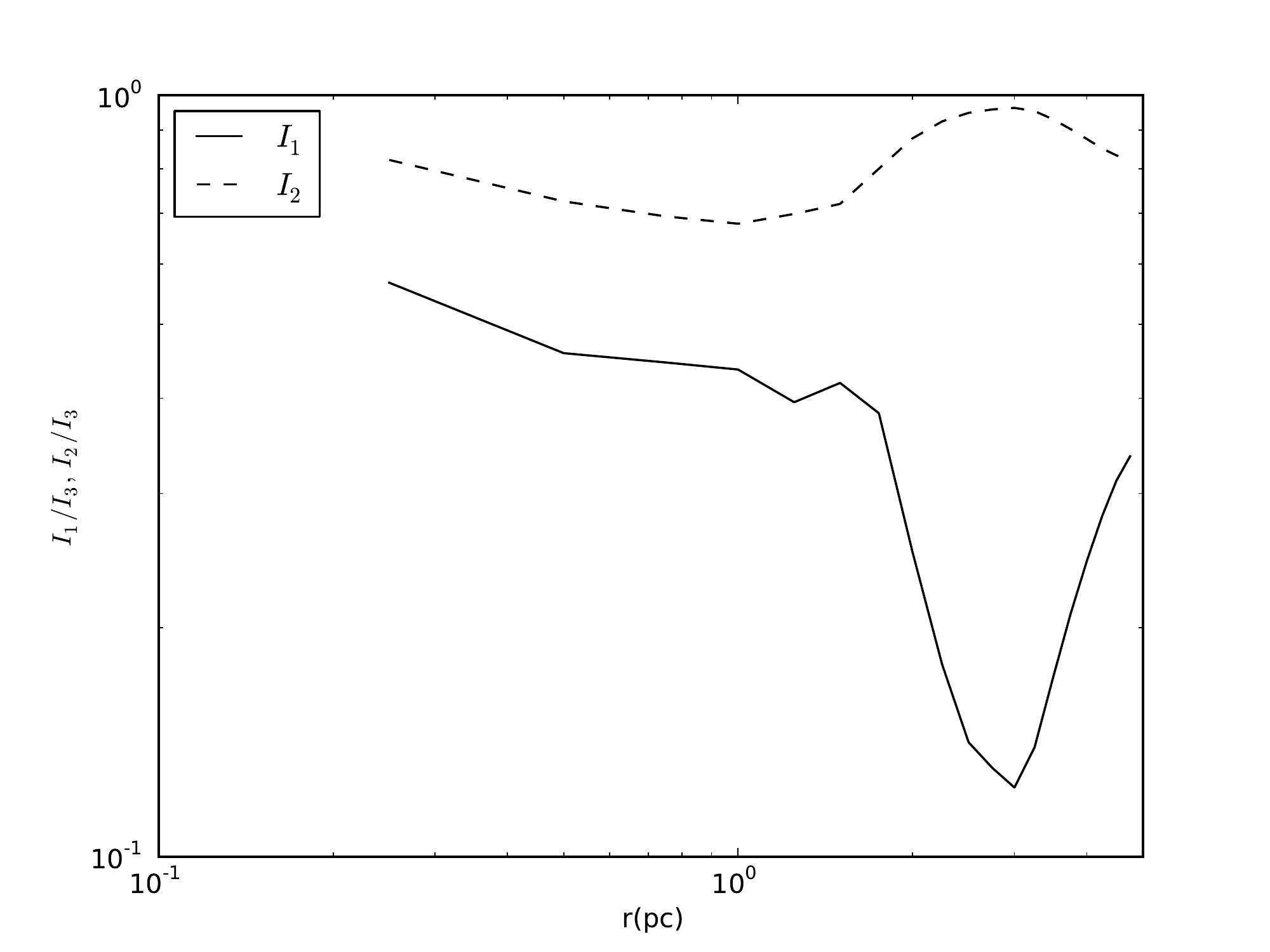}
\caption{\label{fig:momin_flash} Radial profile of moment-of-inertia
  eigenvalues in a FLASH HD run $\sim 0.8{\rm Myr}$ after the first
  star particle forms. The two eigenvalues are normalized by the third
  maximum eigenvalue. Inside of $\approx 1$ pc, the three
  moment-of-inertia eigenvalues are similar to one another, showing
  that the geometry is quasi-spherical. For $1\pc\lesssim r\lesssim
  4\pc$ , $I_2 \approx I_3 \gg I_1$, which suggests that geometry here
  is dominated by a filament. At yet larger $r$, the geometry
  approaches that of a sheet or flattened filament.}
\end{figure}

Figure \ref{fig:momin_flash} shows moment-of-inertia eigenvalues $I_1$
and $I_2$, normalized by the third and largest eigenvalue. In the
construction of a moment of inertia matrix, positions are measured
relative to a star particle. If both $I_1$ and $I_2$ are near 1, the
density distribution is spherical; if $I_1\approx 1$ while $I_2$ is
much smaller, the density distribution is filamentary. A flattened
filament or sheet would have $I_2\lesssim I_1\lesssim1$. Figure
\ref{fig:momin_flash} shows both $I_1$ and $I_2$ to be near unity near
the star particle ($r\lesssim 1\pc$) but $I_2$ to decrease to $\approx
0.1$ for $1\pc\lesssim r\lesssim4\pc$, suggesting that the inner density
structure is nearly spherical while the outer density structure is filamentary. There is a hint that $I_1$ again approaches $I_2$ for
$r>4\pc$, possibly reflecting a sheet-like geometry. This confirms the
visual impression given by Figure \ref{fig:density picture}.

\subsection{Dependence on $\alpha_{\rm vir}$}
\label{ssec:alphavir}

Proposed models of star formation predict $\eff$ to decline with
increasing $\alpha_{\rm
  vir}$~\citep[e.g.,][]{2005ApJ...630..250K,2011ApJ...743L..29H,2012ApJ...759L..27P}. This
is not surprising, since $\alpha_{\rm vir}$ parameterizes the ratio 
between kinetic energy and gravitational potential energy; we expect
objects with low $\alpha_{\rm vir}$ to be bound (and to make stars),
while we expect those with high $\alpha_{\rm vir}$ to be unbound and hence to
make few stars.

This behavior has been verified in numerical simulations
\citep[e.g.,][]{2011ApJ...730...40P,2012ApJ...761..156F}, but exactly
how the mass evolution of individual star particles changes with
$\alpha_{\rm vir}$ has not been studied in detail. 

\begin{figure}
\plotone{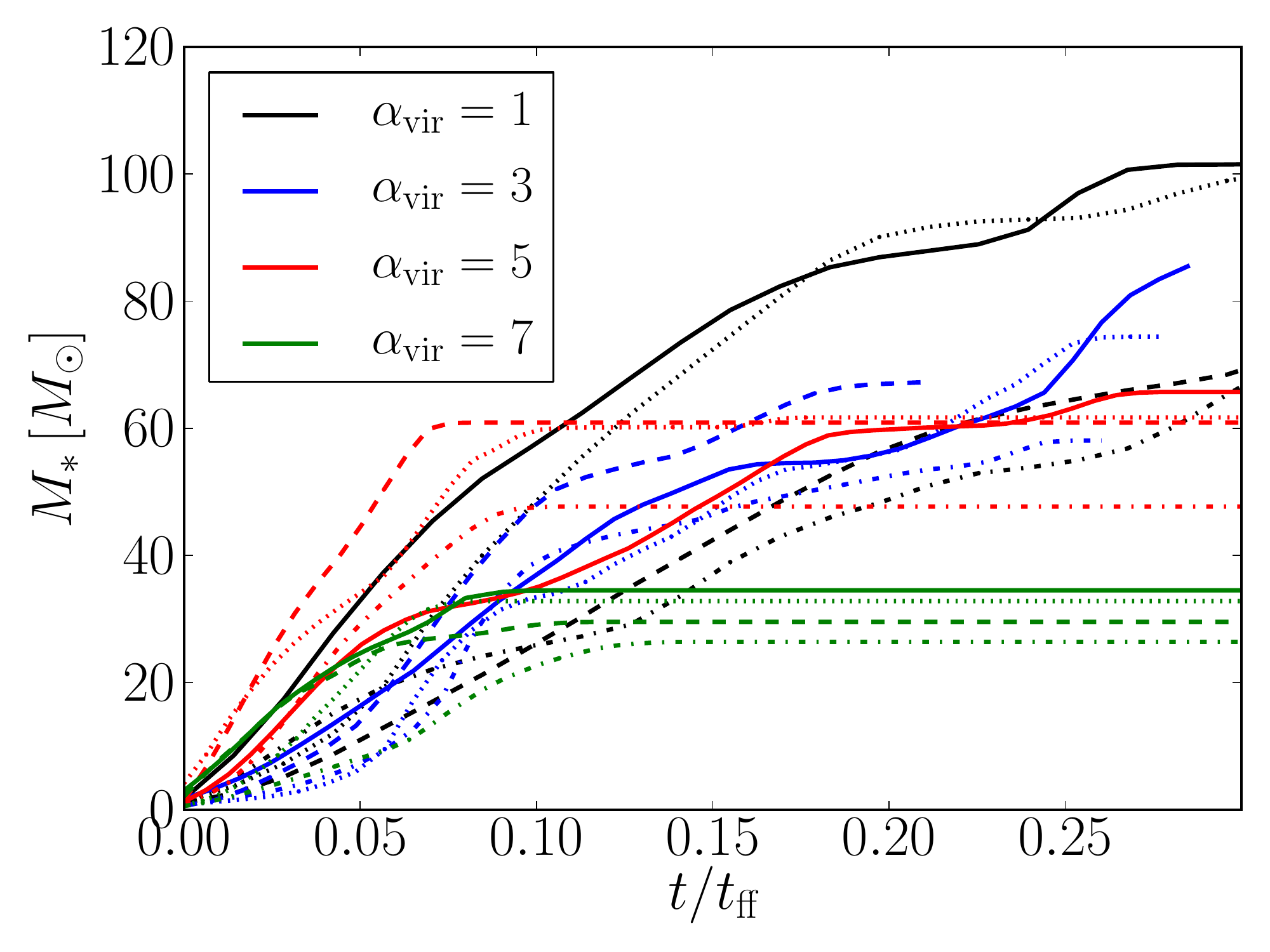}
\caption{\label{fig: alpha_vir cutoff} Stellar mass $M_*$ as a
  function of $t/\tff$ for four runs with different virial parameters:
  $\alpha_{\rm vir}=1$ (black lines), $3$ (blue lines), $5$ (red
  lines), and $7$ (green lines). In this figure $t=0$ corresponds to
  the time at which a given star forms. The life histories of the four most
  massive stars are plotted as solid, dotted, dashed, and dot-dashed
  lines for each value of the virial parameter. At small $t/\tff$, the
  curves all show the characteristic power-law behavior seen in
  Figures \ref{fig:sfe} and \ref{fig:sfe_prescription}. At
  $t/\tff\approx0.05$--0.10 the red and green curves ($\alpha_{\rm
    vir}=5$ and $7$) level off, signaling the cessation of
  accretion. The blue and black curves ($\alpha_{\rm vir}=3$ and $1$)
  continue to rise beyond $t/\tff\approx 0.2$, showing that accretion
  persists for much longer times.}
\end{figure}

Figure \ref{fig: alpha_vir cutoff} shows the accretion histories of
the four most massive stars in each of four runs with different values
of the virial parameter. The figure shows that increasing $\alpha_{\rm
  vir}$ tends to lead to lower mass accretion rates (with substantial
scatter). However, the figure also demonstrates that the accretion is
shut off before $\tff$ is reached when $\alpha_{\rm vir}$ is
large---the green curves ($\alpha_{\rm vir}=7$) become horizontal at
$t/\tff\approx0.1$, while the black curves ($\alpha_{\rm vir}=1$)
continue to accrete beyond $t/\tff=0.25$. Note that both the mass and
the time at which accretion halts are similar between all four star
particles in the same $\alpha_{\rm vir}$ run. In supervirial clouds,
$\tau_{\rm dyn} < \tff$. Because large-scale flows reconfigure the
density structure on the dynamical timescale, in high $\alpha_{\rm
  vir}$ clouds there is less time for star particles to accrete mass
than in low $\alpha_{\rm vir}$ clouds. The globally slower SFR at 
high $\alpha_{\rm vir}$ can then be explained by
both the slower accretion rate onto individual star particles and the
fact that the supply of gas streaming from large radius toward the
star is interrupted when the large-scale turbulent flow varies on a
timescale substantially shorter than the (mean density) free-fall
time.

\section{DISCUSSION}
\label{sec:discussion}

Papers on the SFR in 3D simulations have generally
discussed $\eff$ as a constant value
\citep[e.g.,][]{2011ApJ...730...40P,2012MNRAS.419.3115B,2012ApJ...754...71K,2012ApJ...761..156F,2014MNRAS.439.3420M}. 
However, examination of the figures in these papers shows that the
stellar mass grows in a nonlinear manner. The figures are consistent with our finding
that the SFR rapidly increases with time owing to the
effects of self-gravity.

The time-dependence of the SFR has been recognized by
\citet{2014MNRAS.439.3420M} in their simulations, both HD and
MHD with stellar feedback. Those authors
speculate that the dependency should disappear if turbulence is driven
during the gravitational collapse. Our simulations refute this hypothesis: 
we find $\eff$ to be
time-varying in our gravito-turbulent simulations with 
continuously driven turbulence. \citet{2014MNRAS.439.3420M} report a
global $M_*(t) \sim t^{\alpha_p}$ with $\alpha_p\approx 3$, while
at the same time the same measurement on the most massive stars
results in $\alpha_p=2$ in their HD runs. This is likely because their
stellar population has not yet approached a steady-state distribution
as shown in their Figure 12. 

Simulations that take the formation
  of and gas accretion onto molecular clouds into account via 
	collision of two WNM flows 
have also reported SFR to rise with time 
\citep[e.g.,][]{vazquez09} even with the inclusion of stellar feedback
\citep{vazquez10, colin13} or magnetic field \citep{vazquez11}. 
Our results and interpretations resonate with the works of \citet{cho11} and 
\citet{2012ApJ...750...13C}, who both suggest that gas self-gravity plays an 
important role in determining turbulent statistical properties. 
Through direct comparisons between runs with different kinds of gravitational 
interactions, as well as close examinations of velocity profiles, we have shown 
quantitatively that gas self-gravity accelerates star formation and enhances 
turbulent motion around local density peaks.

Turbulent core-collapse models also predict superlinear stellar mass
growth (e.g., $M_* \propto t^4$ in \citealt{1997ApJ...476..750M} and
$M_* \propto t^2$ in \citealt{2003ApJ...585..850M}). While our SFE
evolution appears to agree well with the model of
\citet{2003ApJ...585..850M}, the turbulent core-collapse model assumes
that gas is in hydrostatic equilibrium (HSE). Our density and velocity
profiles show that the assumption of HSE is not satisfied in our
simulation. What turbulent core-collapse models overlook is that the
nature of turbulence is altered by self-gravity; for example the
power-law index $p$ of the turbulence is altered, as shown in Figure
\ref{fig:velocities_flash}. Using the adiabatically heated turbulence
model of \citet{2012ApJ...750L..31R}, we will show in Paper II that
properly accounting for the interplay between gravity and turbulence
can explain all the numerical results presented in this paper.

Although the main conclusions of our work are that the SFR and SFE 
are dynamic and time-varying quantities, we
emphasize at the same time that our simulations are compatible with
previous work. Indeed, we can reproduce the ``constant'' $\eff$
calculated by, e.g., \citet{2011ApJ...730...40P},
\citet{2012ApJ...754...71K}, and \citet{2014MNRAS.439.3420M}. The
value of $\eff$ reported by these papers is found by fitting a
straight line to the stellar mass $M_*(t)$ at late times, typically a substantial
fraction of a free-fall time after the first stars have
formed. Following their recipe, we find similarly large $\eff\sim 0.3$.

Both \citet{2012ApJ...754...71K} and
\citet{2014MNRAS.439.3420M} find that stellar feedback does not
significantly alter $\eff$, suggesting that the forms of stellar feedback
they include (protostellar jets, ionized gas pressure, and radiation
pressure, calculated using a flux-limited diffusion approximation) do
not regulate the SFR at the scale of their
simulation ($\sim 0.5 {\rm pc}$).
Even in regions of low-mass star formation where protostellar 
outflows become important, the SFE reduction factor is only $\sim$3 
\citep{matzner00,hansen12,machida13,offner14,federrath14}.

In and of itself, rapid local star formation would suggest that the global 
SFR should be much larger than observed. However, we appeal
to other work suggesting that feedback is what determines the global
SFR, by regulating the amount of gas in
gravitationally bound
GMCs \citep{2005ApJ...630..167T,2011MNRAS.417..950H,2013MNRAS.433.1970F}.

\subsection{Slow Star Formation in Supervirial Gas}
\label{sec: large alpha}

The statement that the SFR is of order the cloud mass
divided by the free-fall time (Equation (\ref{eqn: slow})) holds for 
simulations that have global
virial parameters of order unity or smaller. Simulations with virial
parameters larger than 1 show very different behavior---the SFRs are greatly reduced.

We believe that the low SFRs seen in supervirial
simulations reflect the fact that strong turbulence does
not allow large-scale steep ($k_\rho\gtrsim1.5$) power-law density
structures to form. Such large-scale structures are a prerequisite for
the rapid accretion seen at low $\alpha_{\rm vir}$. We note that, even
in supervirial flows, there are regions of convergent flows in which
small-scale bound clumps can form~\citep[e.g.,][]{1999ApJ...527..285B} and 
where the large-scale gravitational field plays little or no role. This is
essentially how stars form (at greatly reduced efficiency) in our 
no-gas-self-gravity runs.

If the free-fall time is much longer than
the dynamical time, then the newly formed star or cluster
will run out of fresh material in a time $t\sim \tdyn$. To see this,
recall Figure \ref{fig:density picture}, which shows that regions of
high density occupy little volume. In high-$\alpha_{\rm vir}$
flows, large-scale turbulence strips the outer layers of these
enhanced density regions on the dynamical time, limiting the amount of
mass that can be accreted onto stars.

\subsection{Gravitational versus Turbulent Collapse}
\label{sec: turbulent collapse}
\citet{2013ApJ...763...51F} perform simulations similar to ours, but
with a broader range of Mach numbers and virial parameters. They
obtain many of the same results we do, including very high values of
$\eff$. They emphasize a different aspect of the result, focusing on 
how variations in the properties of the turbulence affect the SFR: 
for example, they stress that in simulations with large virial parameters, 
or noncompressive driving, the SFR is low compared to regions where virial 
parameter is small or the driving is compressive. We focus on the time 
evolution of the stellar mass in regions where star formation is proceeding.


We have investigated the relative contribution of turbulence and gas 
self-gravity to the time evolution of stellar mass in Section 
\ref{ssec:sfe_evol}, in particular Figure \ref{fig: no self-gravity}.
We show there that turbulence
alone---whether purely solenoidal or purely compressive---in the 
absence of self-gravity, does not drive the rapid
star formation seen in simulations that include self-gravity. If
high Mach number compressive turbulence were solely responsible for
the rapid star formation seen in self-gravitating simulations, then
turning off the self-gravity of the gas should not affect the SFR. 
Furthermore, Figure \ref{fig:PDF_exclude_no_grav}
shows directly that the power-law density tail disappears in the
absence of self-gravity, suggesting that the local overdense regions
that determine the rate of star formation (not just the
seeds of star formation) are generated by
gravity. When gas self-gravity is turned off (but when the subgrid
star formation routine is still active) star formation does proceed,
but at a rate greatly reduced compared to the case where gas
self-gravity operates. In addition, SFRs in the
no-gas-self-gravity runs do not accelerate with time.

The complementary experiment, where self-gravity operates but there is no
turbulent driving, has been reported on many times in the
literature. For example, \citet{2012ApJ...754...71K} find that if the
gas is initialized with a smooth density distribution 
and some turbulent velocity field,
the time to the formation of the first star is relatively long, but
subsequent star formation is exceedingly rapid. By comparison, if the
initial velocity distribution is turbulent (initially driven but decays 
when gravity is turned on) and the emergent density distrubiton 
is a consequence of initial turbulent driving, the first stars
form earlier, but the subsequent SFR grows less
rapidly than in the no-turbulence case.

We conclude that while strong turbulence does hasten the initial
collapse, 
it is not the main determinant of the growth rate of stellar mass 
once star formation begins.
Turbulence
provides the seeds for local collapse but it is the gas self-gravity
that drives accretion onto these seeds. In other words, it is not
the number of seeds (or the initial mass of these seeds) that
determines 
the evolution of SFRs;
the important determinant is the density
structure (and therefore mass) evolution around these individual seeds.

SFRs calculated by integrating over static lognormal
density PDFs from a certain critical density
\citep[e.g.,][]{2005ApJ...630..250K} will, therefore,
underestimate the true rates. Models of star formation need to
properly take into account the dynamical evolution of density PDFs
produced by large-scale effects of gravity.

\subsection{Collapse on Subgrid, Global, and Intermediate Scales}
We distinguish four different scales in our simulations; loss of
support against gravity can occur on all four. The smallest scale,
which we refer to as subgrid, corresponds to a few to several cell
lengths---in our simulations this corresponds to $l\lesssim
0.1\pc$. The loss of support on this and smaller scales is modeled by
a star particle creation routine. 

The global scale is the size of the box in simulations. In
galaxies, we identify the global scale as the local disk scale height,
similar to the sizes of the largest GMCs, or, in low-mass star
formation regions, the size of the host GMC (which can be
substantially below the disk scale height).

We do not see a strong global collapse in any of our turbulent
simulations; this is not surprising, since we turn on self-gravity
only after establishing fairly strong turbulent motions, with
$\alpha_{\rm vir}$ no smaller than one, and then run for a time of
order the free-fall time. In simulations with smooth initial density distributions,
however, global collapse is observed \citep[e.g.,][]{2012ApJ...754...71K}.

Support can also be lost on the local scale whose value depends
on the global virial parameter. In most of our simulations, the local scale spans 
roughly the two decades between the box size and the subgrid
scale:
\be
\Delta x\lesssim l_{\rm local}\lesssim L.
\ee
In our simulations, it is this local scale that is relevant for
setting the pace of star formation. We have shown that the power-law
tail of the density PDF forms on this scale; here the gravitational
potential energy liberated from collapse is converted to turbulent
energy and flattens the local size-linewidth relation.

For larger values of $\alpha_{\rm vir}$, the local scale is
truncated on the high end by large-scale turbulence, reducing the global
SFR dramatically. This `supervirial' truncation scale is 
the fourth scale we identify.

We have shown that rapid star formation occurs before, and even in the
absence of, a global collapse. This result is important: while our
simulations lack feedback, we have argued elsewhere that feedback from
stars prevents global collapse. That same feedback is likely to cut
short the rapid star formation we find here, so that on GMC scales, at
least, the fraction of gas turned into stars is well below the $\sim
50\%$ or higher levels we find here. 

\subsection{Comparison to Observations}

\citet{2007ApJ...654..304K} and \citet{2012ApJ...745...69K} argue from
observations that the SFR per free-fall time on all
scales, including scales at or below that of GMCs, is nearly constant
and equal to $2\%$, an order of magnitude below the rates found in
numerical simulations.

Most of the GMCs they consider are local clouds lacking massive stars
\citep{2010ApJ...723.1019H,2010ApJ...724..687L}. \citet{2011ApJ...729..133M},
in contrast, finds $\eff$ to range from 0.001 to 0.5 in clouds
harboring massive star clusters. \citet{2013MNRAS.436.3167F}
  finds a similar 
scatter in $\eff$ using more observations of nearby clouds and clumps. 
In order to definitively test the constancy of $\eff$ in GMC observations, 
one must consider all clouds, near or far, actively star-forming or not. 
This more complete census is what we
present in Papers III and IV. We find upward of
three orders of magnitude dispersion in $\eff$ and show that our
theoretically derived $M_*\propto t^2$ relation fits the observational
data well. 

The upper end of the observed range in $\eff$ is consistent with the
numerical results found here, and the large dispersion seen in $\eff$
likely reflects cloud-to-cloud variations in age and $\alpha_{\rm
  vir}$. Our results suggest that supervirial clouds will have low
$\eff$, in agreement with \citet{2012ApJ...761..156F}.
We test this suggestion observationally in Paper III.

Direct evidence for time-varying $\eff$ was presented by \citet{palla00}; 
they noted a steeply rising population of younger stars (age $\sim$ 1 Myr) 
compared to older stars (age $\sim$ 10 Myr) in nearby clusters, 
interpreting the trend as accelerating star formation. The age distribution in 
these clusters came into question, however, since sources of error such as 
differential extinction, stellar photometric variation, and contamination from 
field stars can also mimic the spread in stellar ages (\citealt{hartmann01}, 
\citealt{lada03}, and \citealt{hartmann03} but see \citealt{zamora12}).
Other evidence comes from observations of GMCs in the Large Magellanic Cloud. 
\citet{kawamura09} find twice as many clouds with only HII regions 
than those with star clusters, suggesting that clouds spend more time in the early 
stages of star formation.

\section{CONCLUSIONS}
\label{sec:concl}
We have argued that the SFR in gravitationally bound
objects is controlled by gravity. Examples of such objects include the
most massive giant molecular clouds, clumps, and cores. From our
finding that stellar mass grows superlinearly with time, most star
formation will happen over the 1/10 or 2/10 of a
free-fall time. We found that the radial density profile steepens
around density peaks at which stars form; at the same time, a power-law 
tail develops on the high-density end of the lognormal density
PDF. Gas self-gravity is
responsible for all these changes to the density structure. Gas
self-gravity also affects the velocity structure of turbulence, as
shown by velocity enhancements near density peaks. In the absence of
self-gravity, converging flows also effect changes to the density
and the turbulent velocity structures but to a lesser degree.

Unlike the high SFEs seen in our simulations
of virialized clouds, supervirial clouds show much slower growth in
stellar mass; we have argued that this is  because there is not enough
time for gas to collapse before turbulence completely alters the
density field. Our results show that self-gravity acting on scales
larger than 0.1 pc has a direct consequence on cloud density
structure, turbulence, and the rate of star formation. From our
simulation results as well as the agreement with recent observations,
we conclude that SFR is a dynamic, time-varying
property, not a constant as previously thought.

\acknowledgments
We would like to give special thanks to Eugene Chiang for his thorough
proofreading. We thank the anonymous referee, David Collins, Christoph 
Federrath, Stella Offner, and Enrique V\'{a}zquez-Semadeni for helpful comments. 
EJL would like to thank the ENZO development team
for their help with ENZO, and the yt development team \citep{yt} for
their help with the analysis of our simulation data. This research was
supported by Natural Sciences and Engineering Research Council of
Canada through CGS M and PGS D3 scholarship to EJL and Canada
Research Chair program to NM. This work was supported in part
by the National Science Foundation under Grant No. PHYS-1066293 and
the hospitality of the Aspen Center for Physics. PC acknowledges
support from from the NASA ATP program through NASA grant NNX13AH43G,
and NSF grant AST-1255469. Some of the computations were performed on
the gpc supercomputer at the SciNet HPC Consortium
\citep{2010JPhCS.256a2026L}. SciNet is funded by: the Canada
Foundation for Innovation under the auspices of Compute Canada; the
Government of Ontario; Ontario Research Fund - Research Excellence;
and the University of Toronto. The authors acknowledge the Texas
Advanced Computing Center (TACC) at The University of Texas at Austin
for providing HPC resources that have contributed to the research results 
reported within this paper. URL: \url{http://www.tacc.utexas.edu}

\bibliography{accretion}

\end{document}